\documentclass[twocolumn]{aastex6}
\usepackage{times,epsfig,amssymb,amsmath}
\usepackage{graphicx,subfigure}
\usepackage{color}
\usepackage{xparse}
\usepackage{tablefootnote}

\fullcollaborationName{The Friends of AASTeX Collaboration}

\newcommand{\Planck}{{\em Planck}}
\newcommand{\XMM}{{\em XMM}-{\em Newton}}

\newcommand{\mMpl}{M^\text{Pl}_{500}}

\begin{document}


\title{ Spectroscopic confirmation and velocity dispersions for twenty {\em PLANCK} galaxy clusters at $0.16 < z < 0.78$ } 


\author{Stefania Amodeo\altaffilmark{1},
Simona Mei\altaffilmark{1,2,3,4}, 
Spencer A. Stanford\altaffilmark{5},
Charles R. Lawrence\altaffilmark{3},
James G. Bartlett\altaffilmark{3,6},
Daniel Stern\altaffilmark{3},
Ranga-Ram Chary\altaffilmark{8},
Hyunjin Shim\altaffilmark{9},
Francine R. Marleau\altaffilmark{10,7},
Jean-Baptiste Melin\altaffilmark{11},
Carmen Rodr\'iguez-Gonz\'alvez\altaffilmark{8}
}
\altaffiltext{1}{LERMA, Observatoire de Paris, PSL Research University, CNRS, Sorbonne Universit\'es, UPMC Univ. Paris 06, F-75014 Paris, France} 
\altaffiltext{2}{Universit\'{e} Paris Denis Diderot, Universit\'e Paris Sorbonne Cit\'e, 75205 Paris Cedex 13, France} 
\altaffiltext{3}{Jet Propulsion Laboratory, California Institute of Technology, 4800 Oak Grove Drive, Pasadena, California 91011, USA} 
\altaffiltext{4}{Cahill Center for Astronomy \& Astrophysics, California Institute of Technology, Pasadena, CA 91125, USA} 
\altaffiltext{5}{Department of Physics, University of California Davis, One Shields Avenue, Davis, CA 95616, USA ; Institute of Geophysics and Planetary Physics, Lawrence Livermore National Laboratory, Livermore, CA 94550, USA} 
\altaffiltext{6}{APC, AstroParticule et Cosmologie, Universit\'e Paris Diderot, CNRS/IN2P3, CEA/lrfu, Observatoire de Paris, Sorbonne Paris Cit\'e, 10 rue Alice Domon et L\'eonie Duquet, 75205, Paris Cedex 13, France} 
\altaffiltext{7}{National Optical Astronomy Observatory, Tucson, AZ 85719, USA} 
\altaffiltext{8} {Infrared Processing and Analysis Center, California Institute of Technology, Pasadena, CA 91125, USA} 
\altaffiltext{9}{Department of Earth Science Education, Kyungpook National University, Republic of Korea} 
\altaffiltext{10}{Institute of Astro and Particle Physics, University of Innsbruck, 6020, Innsbruck, Austria} 
\altaffiltext{11}{IRFU, CEA, Universit{\'e} Paris-Saclay, F-91191 Gif-sur-Yvette, France} 




\begin{abstract}
We present Gemini and Keck spectroscopic redshifts and velocity dispersions for twenty clusters detected via the Sunyaev-Zel'dovich (SZ) effect by the \Planck\ space mission, with estimated masses in the range $2.3 \times 10^{14} M_{\odot} < \mMpl < 9.4 \times 10^{14} M_{\odot}$.  Cluster members were selected for spectroscopic follow-up with Palomar, Gemini and Keck optical and (in some cases) infrared imaging. Seven cluster redshifts were measured for the first time with this observing campaign, including one of the most distant \Planck\ clusters confirmed to date, at $z=0.782\pm0.010$, PSZ2 G085.95+25.23. The spectroscopic redshift catalogs of members of each confirmed cluster are included as on-line tables. We show the galaxy redshift distributions and measure the cluster velocity dispersions. The cluster velocity dispersions obtained in this paper were used in a companion paper to measure the  \Planck\  mass bias and to constrain the cluster velocity bias.
\end{abstract}




\section{Introduction} \label{sec:intro}
Massive galaxy clusters are sensitive cosmological probes \citep[e.g.][]{allen+11}, yet these are rare objects best found in all-sky surveys covering large volumes. 
The {\em ROSAT} All-Sky Survey \citep[RASS,][]{truemper1993} dates back to the early 1990s and has served the community as a workhorse since, providing hundreds of cluster candidates. A subsequent important step has been taken by the \Planck\ satellite, launched on 2009 May 14. \Planck\ detects clusters based on the Sunyaev-Zel'dovich (SZ) effect \citep{sz1970, birkinshaw1999, carlstrom+2002}, i.e. the distortion of the energy spectrum of cosmic microwave background (CMB) photons passing through the cluster due to inverse Compton scattering with hot electrons. Being independent of distance, the SZ signal does not suffer from cosmological dimming and it is proportional to the cluster mass. Benefiting from this, \Planck\ extends the (X-ray) RASS catalog to higher redshift and contains a large fraction of massive objects of the type most prized for cosmological studies. 

\Planck\ has produced two all-sky cluster surveys through the SZ effect \citep{Planck2014, Planck2015}: the PSZ1 with 1227 candidates based on 15.5 months of data, and the PSZ2 with 1653 candidates from the full mission dataset of 29 months.  Of the PSZ2 candidates, 1203 have been confirmed by ancillary data and 1094 have redshift estimates, in the range $0<z<1$, with a mean redshift of $z\sim0.25$. The mean mass of the confirmed clusters over the whole redshift range is $\mMpl = 4.82\times10^{14} M_\odot $ 
(see the definition of $\mMpl$ below).  

The \Planck\ collaboration has undertaken a large follow-up effort to confirm cluster candidates and measure their redshifts.
The first optical follow-up was based on observations with the Russian-Turkish 1.5 m telescope \citep{planck_int_xxvi} and provided spectroscopic redshifts of 65 \Planck\ clusters.  
The second optical follow-up, based on observations with telescopes at the Canary Islands Observatories, yielded 53 cluster spectroscopic redshifts \citep{planck_int_xxxvi}.
The \Planck\ collaboration has also carried out X-ray validation programs with \XMM\ \citep{planck_early_ix,planck_int_i,planck_int_iv}, where redshifts for 51 clusters were obtained from X-ray spectral fitting.

Our follow-up program presented in this paper  includes the spectroscopic follow-up of 20 \Planck\ cluster candidates with the Gemini and Keck telescopes (P.I. J.G. Bartlett and F.A. Harrison, respectively). The goals of our programs were: (1) to confirm \Planck\ SZ detections as clusters and measure their redshifts; (2) to estimate their masses using cluster galaxy velocity dispersions; and (3) to measure the  \Planck\  mass and velocity bias. We use SDSS \citep[Sloan Digital Sky Survey;][]{york+00}, and Palomar and Gemini imaging to select the cluster galaxies to target with spectroscopy. 

In this paper, we describe our observations and publish the optical spectroscopy of cluster members, from which we derive the cluster redshifts and velocity dispersions.
In a companion paper \citep{amodeo+17}, we use these observations to estimate the clusters' dynamical masses and calibrate the all-important relation between the SZ Compton parameter, $Y$, and mass.

The paper is organized as follows. In Section \ref{sec:data} we present our sample of \Planck\--selected clusters and describe the observing programs carried out at the Palomar, Gemini and Keck telescopes. 
A table describing all the targets observed with the Palomar telescope is given in Appendix \ref{sec:palomar}. 
In Section \ref{sec:red} we describe the spectroscopic redshift and galaxy velocity dispersion measurements. 
For clusters with a spectroscopic follow-up, we include figures of redshift histograms, optical images and SZ maps in Appendix \ref{sec:figures}.
Catalogs of cluster member galaxies with spectroscopic measurements are included as on-line tables. We illustrate the parameters published in the catalogs in Section \ref{sec:catalog} and give an example in Appendix \ref{sec:appendix}.
In Section \ref{sec:discussion} we discuss our results in the context of optical identifications of \Planck\ clusters. 
 
Throughout this paper, masses are quoted at a radius $R_\Delta$, within which the cluster density is $\Delta$ times the critical density of the universe at the cluster's redshift, where $\Delta=\{200, 500\}$. 
We refer to the $\Delta=200$ radius as the ``virial radius", $R_\text{200}$. Mass and radius are directly connected via $M_\Delta \equiv \Delta \,{H_z}^2  {R_\Delta}^3 /(2 G)$, where $H_z$ is the Hubble constant at the cluster's redshift.

\section{Data and Observations} \label{sec:data}

In this section we describe our spectroscopic observations with the Gemini and Keck telescopes, and the Palomar telescope imaging that was used to select cluster members. The details of each observing run (pre-imaging and optical spectroscopy) are listed in Table \ref{obs}.

Since it is well-known that early-type galaxies (ETGs) in clusters define a tight red-sequence up to redshift $z\sim1.5$ \citep{mei+09}, and can be easily identified with respect to field background galaxies, we selected cluster members to follow-up with spectroscopy from optical and infrared imaging using a red sequence search method \citep{Gladders2000,licitra+16a,licitra+16b}. For most clusters, we used $g'$ and $i'$ filters for imaging, since the ETG $(g-i)$ color is monotonic over the redshift range in which most of \Planck\ clusters are detected, $z<1$. We also observed the $r'$ band, when possible within our exposure time constraints, to obtain better photometric redshifts. For the candidates that appeared to be at $z>0.6$ from their {\em WISE} imaging in the mid-infrared \citep[see the {\em WISE} analysis in][]{Planck2015}, we obtained near-infrared observations in the $J$ and $K$ bandpasses. For some of our targets, we could not obtain images at two different wavelengths and used SDSS photometry when available.  

Cluster members were selected as red sequence galaxies by their colors, using \citet{BC2003} stellar population models and \citet{mei+09} empirical red sequence measurements, following the cluster member selection technique described in \citet{licitra+16a,licitra+16b}, adapted for the bandpasses available for these observations. 

\begin{table*}[ht!]
\begin{center}
\caption{Observation details. \label{obs}}
\vspace*{0.2cm}
\begin{tabular}{c c c c c c}
\tableline \tableline
Run & Semester &PI&Tel./Inst.&Program ID&$N_{\rm cl}$\\
\tableline \tableline 
1&2010B&Lawrence&Palomar/LFC,WIRC&&11\\
2&2011A&Lawrence&Palomar/LFC&&25\\
3&2011B&Lawrence&Palomar/LFC&&15\\
2&2011A&Bartlett&Gemini-N/GMOS&GN-2011A-Q-119&11\\
3&2011B&Bartlett&Gemini-N/GMOS&GN-2011B-Q-41&11\\
4&2012B&Lawrence&Palomar/LFC&&9\\
5&2012A&Bartlett&Gemini-S/GMOS&GS-2012A-Q-77&9\\
6&2013B&Harrison&Keck/LRIS&UT 2013 October 4-5&1\\
\tableline \tableline \\
\end{tabular}
\end{center}
\end{table*}

\subsection{Gemini Observations}

The Gemini imaging and spectroscopic follow-up was performed  with GMOS-N and GMOS-S at the Gemini-North and Gemini-South Telescopes, respectively, in the programs GN-2011A-Q-119, GN-2011B-Q-41, and GS-2012A-Q-77 (P.I. J.G. Bartlett). This sample consists of 19 \Planck\--detected galaxy clusters, 17 of which are part of the \Planck\ PSZ2 catalog \citep{Planck2015}, and one is published in the \XMM\ validation follow-up of \Planck\ cluster candidates \citep{planck_int_iv}. Two clusters are not part of the already published \Planck\ papers: (1) PLCK G183.33-36.69 has a detection signal-to-noise ratio ($S/N$)  just below the  \Planck\  catalog selection threshold and (2) PLCK G147.32-16.59 is in the  \Planck\  cluster mask. 

\begin{figure*}[!ht]
\begin{center}
\figurenum{1}
\epsscale{0.5}
\plotone{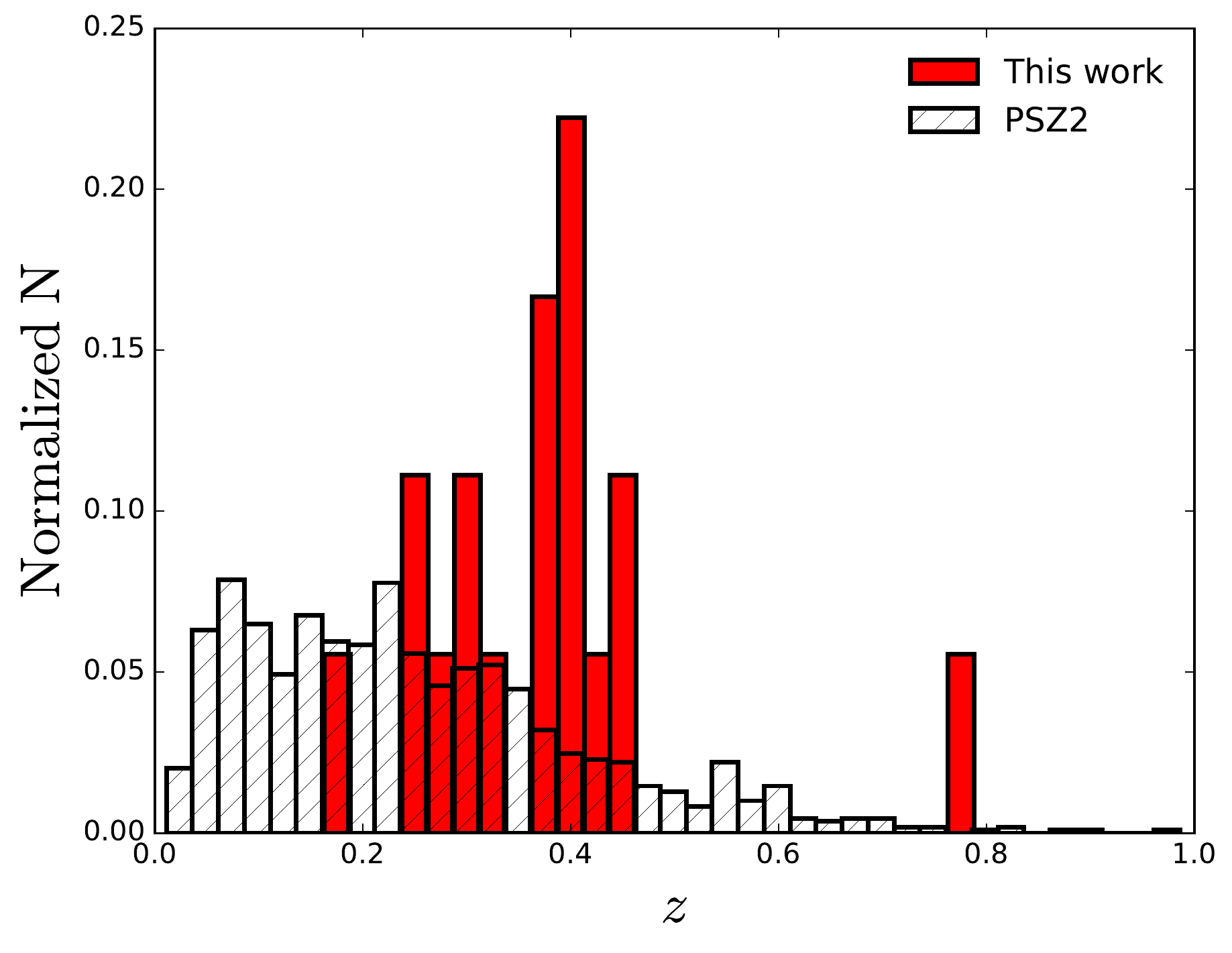}
\plotone{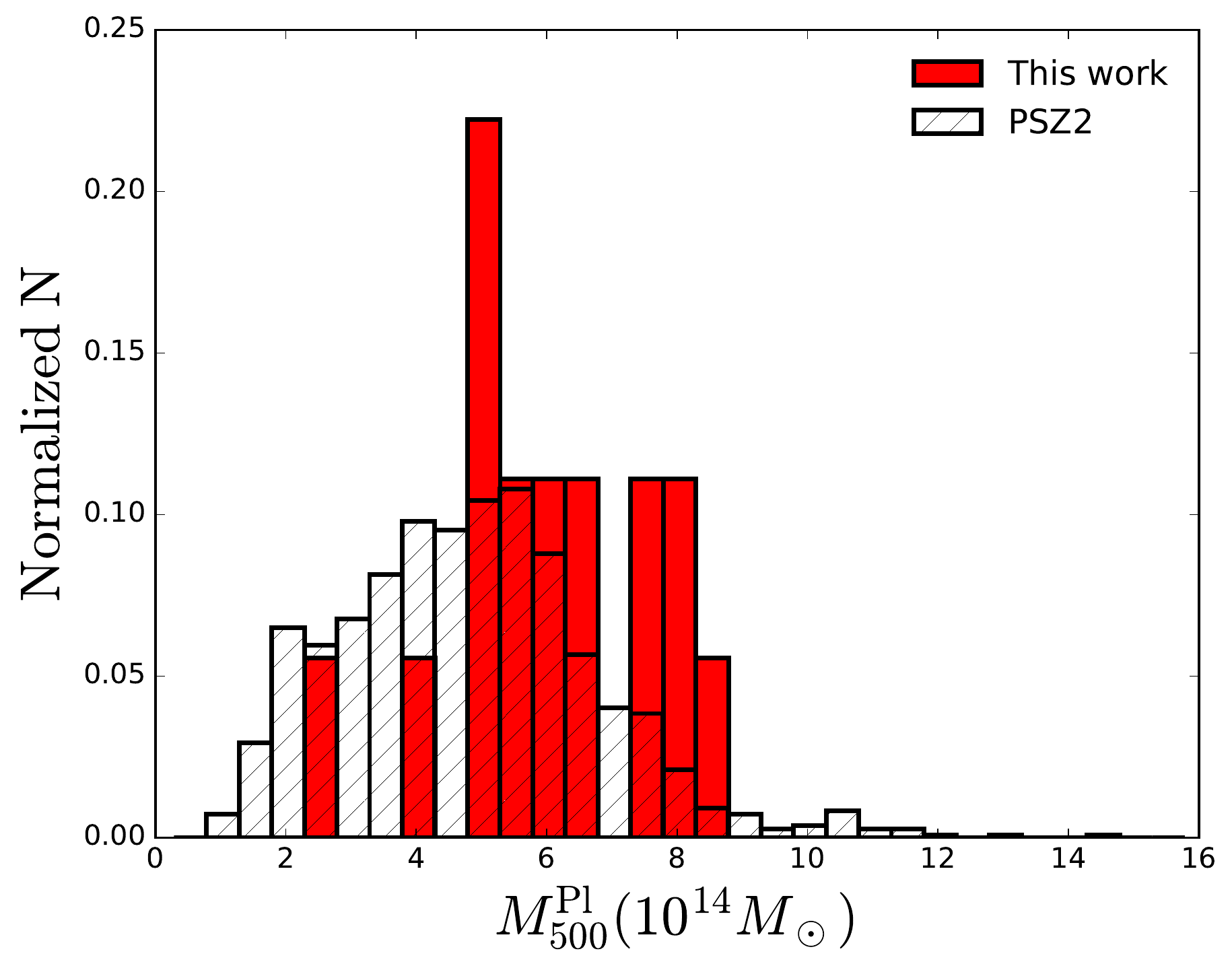}
\caption{Histograms of the redshifts (left) and the masses (right) of our spectroscopic sample compared to the full PSZ2 catalog. These histograms are normalized to the total number of objects in each sample. We have selected cluster candidates with redshift $z>0.2$ (with average redshift larger than the PSZ2 catalog), and larger average mass than the PSZ2 catalog, with cluster masses in the range $2.3 \times 10^{14} M_{\odot} < \mMpl < 9.4 \times 10^{14} M_{\odot}$.  The cluster mass shown here is the \Planck\ mass proxy \citep{Planck2015}.}
\label{fig:hist}
\end{center}
\end{figure*}

The goal of our Gemini program was to obtain a statistical calibration of the \Planck\ SZ mass estimator. For this purpose, we mostly chose clusters that were detected with a  \Planck\  SZ $S/N$ of about 4.5~$\sigma$ or larger, distributed in the Northern and Southern Hemispheres, spanning a wide range in \Planck\ SZ masses, $2\times 10^{14} M_{\sun} \lesssim \mMpl \lesssim 10^{15} M_{\sun}$, in the redshift range $0.16<z<0.44$. In Figure \ref{fig:hist}, we compare our sample to the full PSZ2 catalog. These histograms show that our selection has  an average redshift larger than the PSZ2 catalog, and a mass range covering most of the mass range of the PSZ2 catalog. In fact, our sample has an average redshift of $z=0.37$ and an average mass of $M=6.2\times 10^{14} M_{\sun}$, compared to the average PSZ2 redshift and mass of  $z=0.25$ and $4.8\times 10^{14} M_{\sun}$, respectively. The larger average redshift was chosen to cover most of the cluster members within $\sim R_{200}$ in the field of view of the Gemini and Keck telescopes.

The Northern sample was selected in the area covered by the SDSS, and we used the SDSS public releases and our GMOS-N pre-imaging in the {\it r}-band (150~s) to detect red galaxy over-densities around the \Planck\ detection center.  When unknown, we estimated the approximate cluster redshift using its red sequence to calculate the appropriate exposure times for the spectroscopic follow-up. For PSZ2 G139.62+24.18, PSZ2 G157.43+30.34 and PLCK G183.33-36.69, we used imaging obtained with the Palomar telescope.  For the Southern sample, we obtained GMOS-S pre-imaging in the {\it g} and {\it i}-bands (200~s and 90~s integrations, respectively).

\begin{figure*}[!h]
\begin{center}
\figurenum{2}
\epsscale{0.8}
\plotone{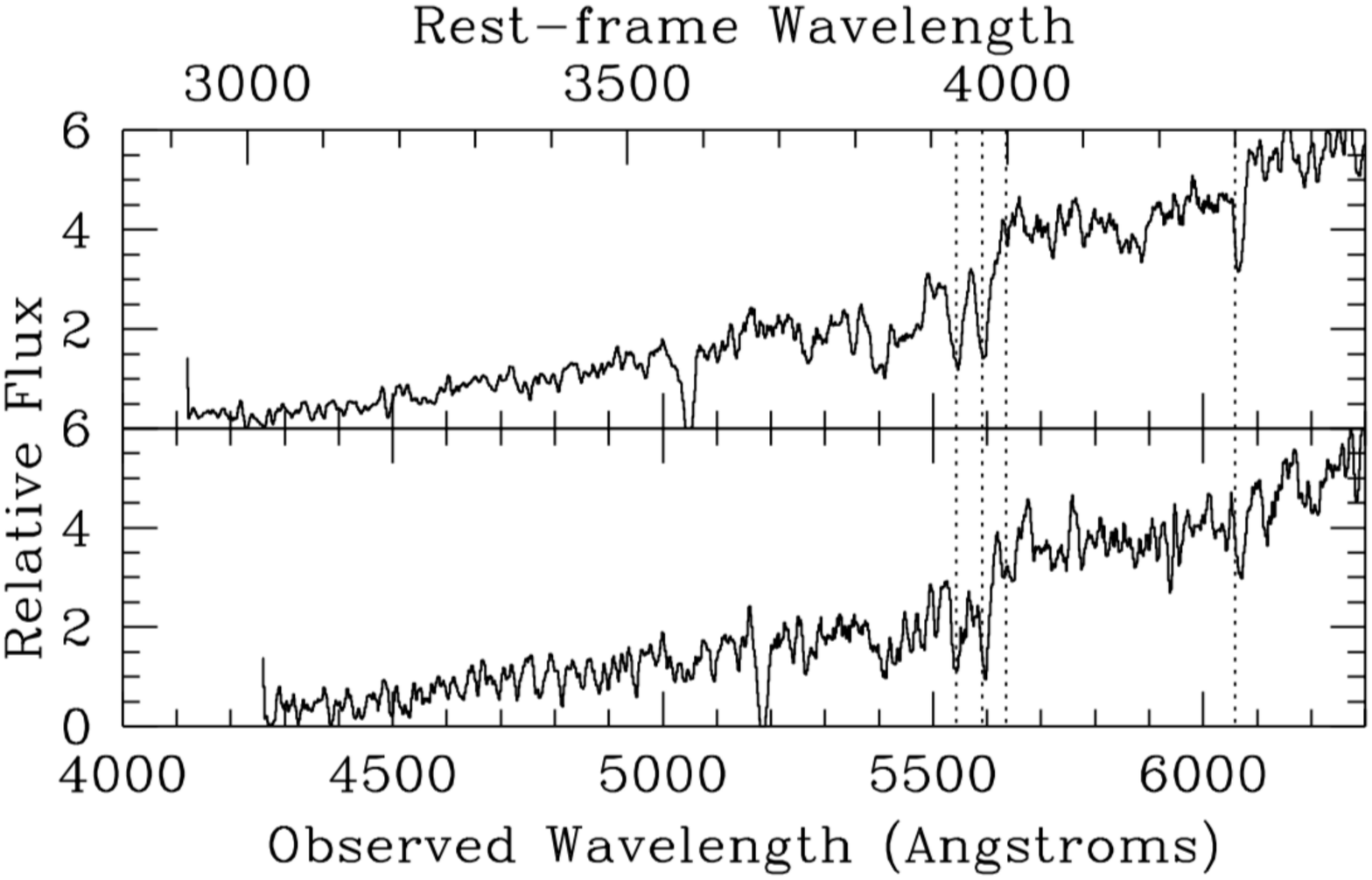}
\caption{Spectra obtained with Gemini/GMOS for two galaxies in the cluster PSZ2 G250.04+24.14 ($z=0.411$). The vertical dotted lines represent Ca H+K, D4000, and the G-band, respectively.}
\label{fig:spectra}
\end{center}
\end{figure*}

Our GMOS spectroscopic observations were reduced using the IRAF Gemini GMOS package and standard  techniques.  After co-adding the reduced exposures, we extracted one-dimensional spectra for the objects in each slitlet and initially inspected them visually to identify optical features such as the 4000 \AA ~break, G-band, Ca H$+$K absorption lines, and, rarely, [O~II]$\lambda$3727 emission. We determined more precise galaxy redshifts by running the IRAF task {\it xcsao}. In Figure \ref{fig:spectra}, we show two Gemini/GMOS spectra of galaxies in the cluster PSZ2 G250.04+24.14.

The clusters that we followed-up with the Gemini telescopes are listed in Table \ref{sample} \citep[see also Table 1 from][]{amodeo+17}. The mass calibration derived from the velocity dispersions of the clusters in this sample is discussed in \citet{amodeo+17}, in which we measured the \Planck\  mass bias and constrained the cluster velocity bias.

\begin{table*}[ht!]
\begin{center}
\caption{Spectroscopically confirmed cluster sample. Clusters are named after their PSZ2 ID, when available. When it is not available, we use the prefix `PLCK' followed by a notation in Galactic coordinates similar to that used in the PSZ2 paper. Right ascension and declination indicate the optical cluster centre. Filter names used for imaging, spectroscopic observing times and the number of masks are also stated. The last column lists the observing run(s) for each target, including pre-imaging.}
\label{sample}
\vspace{0.25cm}
\begin{tabular}{lrrcccccccccccccc}
\tableline \tableline\\
Name & R.A.&Decl.&Filter&$t_{\text{exp}}$&$N_{\text{mask}}$&Run\\
&(deg)&(deg)&&(s)&  &\\
\tableline \tableline 
PSZ2 G033.83-46.57 &326.3015&-18.7159&\it g,i&1800&2&GS-2012A-Q-77\\
PSZ2 G053.44-36.25 &323.8006&-1.0493&\it r&1800&1&GN-2011A-Q-119,GN-2011B-Q-41\\
PSZ2 G056.93-55.08 &340.8359&-9.5890&\it r&1800&2&GN-2011A-Q-119,GN-2011B-Q-41\\
PSZ2 G081.00-50.93 &347.9013&3.6439&\it r&1800&1&GN-2011A-Q-119,GN-2011B-Q-41\\
PSZ2 G083.29-31.03 &337.1406&20.6211&\it r&1800&1&GN-2011A-Q-119,GN-2011B-Q-41\\
PSZ2 G085.95+25.33 &277.6164&56.8823&--&3600&2&Keck Telescope\\
PSZ2 G108.71-47.75 &3.0715 &14.0191&\it r&1800&2&GN-2011A-Q-119,GN-2011B-Q-41\\
PSZ2 G139.62+24.18 $^\text{a}$ &95.4529&74.7014&\it r&900&2&GN-2011A-Q-119,GN-2011B-Q-41\\
PLCK G147.32-16.59 $^\text{b}$ & 44.1101&40.2853&\it r&1800&2&GN-2011A-Q-119,GN-2011B-Q-41\\
PSZ2 G157.43+30.34 $^\text{a}$ &117.2243&59.6974&\it r&3600&2&GN-2011A-Q-119,GN-2011B-Q-41\\
PLCK G183.33-36.69 $^\text{a}$ &57.2461&4.5872&\it r&1800&2&GN-2011A-Q-119,GN-2011B-Q-41\\
PSZ2 G186.99+38.65 &132.5314&36.0717&\it r&1800&2&GN-2011A-Q-119,GN-2011B-Q-41\\
PSZ2 G216.62+47.00 &147.4658&17.1196&\it r&1800&2&GN-2011A-Q-119,GN-2011B-Q-41\\
PSZ2 G235.56+23.29 &134.0251&-7.7207&\it g,i&900&2&GS-2012A-Q-77\\
PSZ2 G250.04+24.14&143.0626&-17.6481&\it g,i&1800&2&GS-2012A-Q-77\\
PSZ2 G251.13-78.15 &24.0779&-34.0014&\it g,i&900&2&GS-2012A-Q-77\\
PSZ2 G272.85+48.79 &173.2938&-9.4812&\it g,i&900&2&GS-2012A-Q-77\\
PSZ2 G329.48-22.67 &278.2527&-65.5555&\it g,i&900&2&GS-2012A-Q-77\\
PSZ2 G348.43-25.50 &291.2293&-49.4483&\it g,i&900&2&GS-2012A-Q-77\\
PSZ2 G352.05-24.01 &290.2320&-45.8430&\it g,i&1200&2&GS-2012A-Q-77\\
\tableline \tableline\\
\end{tabular}
\end{center}
\footnotetext{Also observed at Palomar, see Table \ref{tab:palomar}.}
\footnotetext{Target PLCK G147.32-16.59 is confirmed in the \XMM\ cluster validation \citep{planck_int_iv}, but it is not included in the two \Planck\ catalogues of SZ sources released so far.}
\end{table*}

\subsection{Keck Observations}

We obtained spectroscopy of PSZ2 G085.95+25.23 on the nights of
UT 2013 October 4-5 using the dual-beam Low Resolution Imaging
Spectrometer \citep[LRIS;][]{oke+1995} on the Keck I telescope atop
Mauna Kea.  These slitmask observations were obtained with the 400
$\ell$ mm$^{-1}$ grism on the blue arm of LRIS ($\lambda_{\rm blaze}
= 3400$~\AA), the 400 $\ell$ mm$^{-1}$ grating on the red arm of
LRIS ($\lambda_{\rm blaze} = 8500$~\AA), and the 5600~\AA\ dichroic
was used to split the light.  We obtained three 1200~s integrations
on the first night through variable cloud cover, and two 1200~s
integrations on the second night in photometric conditions.  After
some experimentation, we base our analysis on the single best exposure
from the first night combined with the two exposures from the second
night.  The data were processed using standard techniques within
IRAF, and flux calibrated using standard stars from \citet{massey+gronwall1990} observed on the second night.

\begin{figure*}[!ht]
\begin{center}
\figurenum{3}
\epsscale{0.8}
\plotone{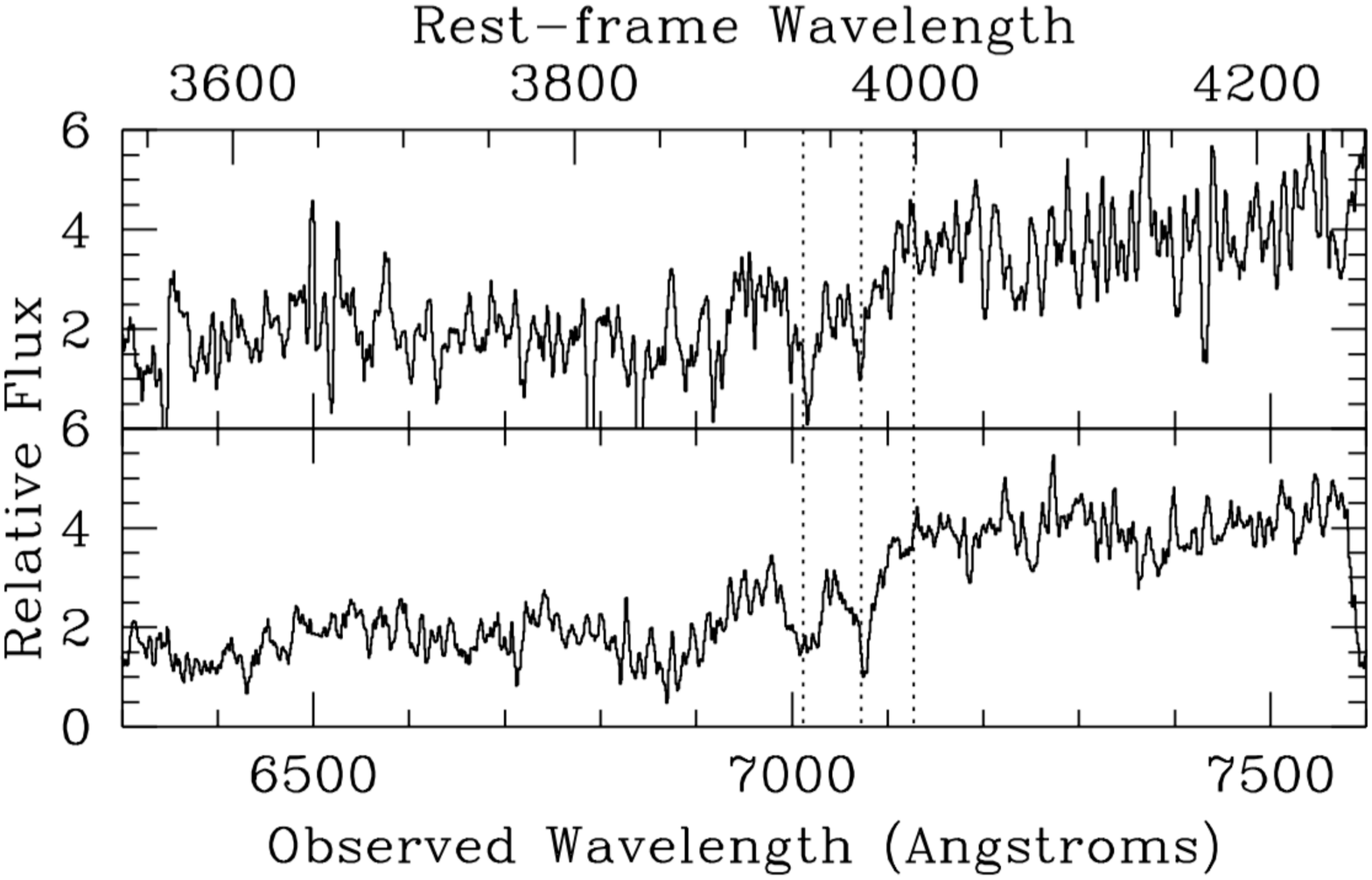}
\caption{Spectra obtained with Keck/LRIS for two galaxies in the cluster PSZ2~G085.95+25.23 ($z=0.782$). The vertical dotted lines represent Ca H+K and D4000, respectively.}
\label{fig:keckspectra}
\end{center}
\end{figure*}

In Figure \ref{fig:keckspectra}, we show two Keck/LRIS spectra of galaxies in the cluster PSZ2~G085.95+25.23.

\subsection{Palomar Optical and Infrared Imaging}

The Palomar optical and infrared imaging, used to select cluster members, was obtained with a dedicated \Planck\ follow-up program  (PI: C. Lawrence) that included several runs. Our Palomar sample is presented in the Appendix, in Table \ref{tab:palomar}.

For the Palomar/WIRC data reduction, we pre-processed the images using a dedicated IRAF package \texttt{noao.imred.ccdred}. 
Master dark frames of different exposure times were constructed 
for each night of observation, 
and these were subtracted from 
science images of the corresponding exposure time. 
Dark-subtracted individual science images 
were then divided by the master flat image in the same filter.  
We tested two ways of constructing the master flat image: 
first, by combining dome flats after the dark correction 
(master domeflat) and, second, by median combining all science images
(master skyflat). Since there was little difference 
between the two master flat images, we chose to use the master domeflat
in the flat correction. Sky subtraction, bad pixel and cosmic ray masking, image aligning 
and combining were done using the IRAF package 
\texttt{xdimsum}\footnote{Experimental Deep Infrared Mosaicing Software}. 
Bad pixel masks were created from the master skyflat image 
by identifying bad, hot, or warm pixels significantly ($>20\,\sigma$) 
lower or higher than the average background. 
After calculating the shifting of the images that were to be sky subtracted, 
we performed the sky subtraction correcting for bad pixels and cosmic rays. The six adjacent images were used to calculate the sky. 
Once the mosaic image was created, we created an object mask from the mosaic image, 
repeated the sky subtraction and combining process  
to obtain the final mosaic image as well as the exposure time map. 
Astrometric calibration was done using the 2MASS point source catalog as a reference 
through the IRAF package \texttt{ccmap}.

The optical Palomar/LFC data reduction comprised basic pre-processing (i.e., bias, dark, flat, crosstalk and 
overscan/trim correction), satellite trail removal, bad pixel
and cosmic ray correction, and aligning and co-adding individual images 
to produce the final mosaic images. Since the LFC data were stored 
in a multi-extension fits (MEF) format, we mostly used 
the IRAF package \texttt{mscred} \citep{valdes+95} as well as 
\texttt{ccdred} for the analysis.  The initial bad pixel masks were generated from the ratio between 
two flat-field images with different exposure times. After the bias, dark,
flat, crosstalk and overscan correction using \texttt{ccdproc}, 
satellite trails in each image were identified using the IRAF task 
\texttt{satzap} and corrected. In addition, initial bad pixels 
and cosmic rays were identified using average sigma clipping. 
The updated bad pixel masks were converted to weight images 
used later in the coadding step.  

Fringe correction was necessary for images taken with the
$i'$-band filter. The fringe effect is less 
noticeable in images with short exposure times ($<300$~s), but 
the interference pattern significantly affects
the background for longer exposures. For fringe correction, we first made an object mask and
the resulting sky map for each image using the IRAF \texttt{nproto.objmasks}.
Then the output sky maps were combined using \texttt{mscred.sflatcombine}
to produce the response sky image, from which the median-filtered response
was subtracted to derive the fringe pattern. Using the fringe pattern 
as the input in \texttt{mscred.rmfringe}, the fringes in $i'$-band
images were successfully divided out. 
After these corrections, astrometric calibration was done with \texttt{ccmap}
using the USNO-B1.0 catalog as a reference. Then the images for each target
were registered and mosaicked using the Terapix/\textit{Swarp} software. 
The images were background subtracted, resampled, and combined 
to produce weighted means of the individual images for flux conservation.  
The weight images previously created from the final bad pixel masks were used. 

Many of our Palomar nights were not photometric, and we could not obtain accurate photometric redshifts with only a few bandpasses. 
However, we could use SDSS and our Palomar images to select cluster member candidates for our Gemini spectroscopic observations.

\section{Cluster Confirmation and Spectroscopic Redshift Measurements} \label{sec:red}

We calculated the cluster redshifts and velocity dispersions using the ROSTAT software \citep{beers+90} with the biweight method (see Table \ref{specres}). This is appropriate to our clusters where there are typically 20 confirmed members. We also report the dispersion $\sigma_\text{G}$ determined from the gapper estimator (as implemented in ROSTAT), which is to be preferred for clusters with fewer than 10-15 members \citep[see][]{girardi+93, girardi+05}. We find that biweight and gapper estimates are perfectly consistent, with the absolute difference between the velocity dispersions calculated from the two methods being on average of ($0.04\pm0.14$) sigma, and never higher then 0.5 sigma.
Since the line-of-sight cluster velocity dispersion can be highly anisotropic, small galaxy samples lead to large systematic uncertainties, with estimated uncertainties of $\lesssim$10\% \citep{white+10} for samples with more than $\sim$ 10-15 galaxies like ours.

We retain as possible cluster members the galaxies within 3$\sigma$ of the average cluster velocity/redshift.  Standard deviations are in the range 0.001-0.008 in redshift, for the clusters that we confirm, apart PLCK G147.32-16.59 that shows evidence for an undergoing merger event (see discussion below). 
Figures \ref{fig:zhist} and \ref{fig:zhist2} show the redshift distributions of the cluster member galaxies (left), the optical image of the cluster with the selected members (middle), and the SZ maps in units of $S/N$ (right), for the Northern and the Southern samples, respectively.  We also present Gaussian fits to the redshift distributions in the left-hand panels.

The middle panels of Figures \ref{fig:zhist} and \ref{fig:zhist2} show the optical pre-imaging, within the Gemini field of view of $5.5\times5.5~\text{arcmin}^2$. Spectroscopically confirmed members are indicated by green circles. 

For PSZ2 G056.93-55.08 we visually observe three spatially separated galaxy groups, but all at the same redshift and within one virial radius.
We derived the virial radius $R_\text{200} = (2.00\pm0.05)$ Mpc from the SZ mass estimate of $\mMpl = (9.4\pm 0.5)\times10^{14} M_\odot$ \footnote{See Appendix A in \citet{amodeo+17} for the conversion from $\mMpl$ to $M^\text{Pl}_{200}$.}. At the cluster redshift, $z = 0.443$, 2 Mpc correspond to 5.7 arcmin in a Planck cosmological model \citep{PlanckCOSMO}. 
We cannot obtain a separate mass estimate for each group because the Planck beam includes all the three groups and we do not have enough spectroscopic members of each group for deriving the group mass from velocity dispersions. Therefore, we consider the three groups as being part of a single cluster detection. 

For all targets but PSZ2 G352.05-24.01, the red circled area is centered on the optical center of the cluster and has a 1 arcmin radius.   The optical center was obtained as the brightest cluster member in the densest cluster region, following a modified version of the centering algorithm from \citet{licitra+16a}. 
For PSZ2 G352.05-24.01, we used the coordinates of the X-ray center, marked with a red cross. 

In the right-hand panels, we show the SZ maps with the same area enclosed by the black circles and centered on the optical position. The SZ maps have an angular resolution of 5 arcmin and are given in units of $S/N$. All the detections lie above $S/N= 4.5$, except for PLCK G183.33-36.69 with $S/N=2$. 

\begin{table*}[ht!]
\begin{center}
\caption{Columns from left to right list the cluster ID, our measured spectroscopic redshift, the new spectroscopic redshift estimates, redshift estimates obtained including the available redshifts in the SDSS DR14,  the total number of galaxies with measured redshifts in the cluster field, the number of confirmed member galaxies, and our measured velocity dispersions using the biweight and the gapper methods \citep{beers+90}. The next three columns give, respectively, the signal-to-noise ratio, the number of detection methods and the \Planck\ mass proxy, as reported in the PSZ2 catalog (we calculated these numbers for the two objects not listed in the PSZ2 catalog).  The last three columns list, respectively, the Kolmogorov-Smirnov (K-S) and the Shapiro-Wilk (S-W) statistics for the probability that the redshift distributions are Gaussian, and the K-S test for a uniform distribution.}
\label{specres}
\vspace{0.25cm}
\resizebox{\textwidth}{!}{
\begin{tabular}{lcccccccccccccc}
\tableline \tableline\\
Name & $z_\text{spec}$ & New $z_\text{spec}$ &$z_\text{spec+DR14}$& $\text{N}_\text{tot}$ & $\text{N}_\text{gal}^\text{conf}$ & $\sigma_\text{BI}$ & $\sigma_\text{BI+DR14}$& $\sigma_\text{G}$ & SNR & Det. Meth. & $M_\text{500}^\text{Pl}$&K-S gaussian prob. &S-W gaussian prob.&K-S uniform prob.\\
&&&&&(km/s)&&&$(10^{14}M_\odot)$&\\
\tableline \tableline
PSZ2 G033.83-46.57 &$0.439\pm0.001$&+&&10&8&$985\substack{+451\\-277}$&&$1051\substack{+309\\-214}$&4.6&2&$5.4\substack{+0.7\\-0.8}$&0.96&0.71&0.50\\
PSZ2 G053.44-36.25 &$0.331\pm0.001$&+&$0.3295\pm0.0003$&21&20&$1011\substack{+242\\-131}$&$1215\substack{+167\\-100}$&$1025\substack{+224\\-117}$&8.9&3&$7.5\substack{+0.5\\-0.6}$&0.99&0.80&0.07\\
PSZ2 G056.93-55.08 &$0.443\pm0.001$&&$0.4430\pm0.0001$&49&46&$1356\substack{+192\\-127}$&$1331\substack{+194\\-128}$&$1345\substack{+170\\-113}$&11.5&3&$9.4\pm0.5$&0.76&0.12&0.01\\
PSZ2 G081.00-50.93 &$0.303\pm0.001$&+&$0.4430\pm0.0001$&15&15&$1292\substack{+360\\-185}$&$1552\substack{+175\\-154}$&$1300\substack{+326\\-140}$&9.2&3&$6.7\pm0.5$&0.97&0.96&0.14\\
PSZ2 G083.29-31.03 &$0.412\pm0.002$&&$0.4123\pm0.0001$&21&20&$1434\substack{+574\\-320}$&$1153\substack{+111\\-94}$&$1591\substack{+376\\-262}$&9.1&3&$7.8\substack{+0.5\\-0.6}$&0.83&0.90&0.004\\
PSZ2 G085.95+25.23 &$0.782\pm0.003$&+&&16&14&$1049\substack{+210\\-180}$&&$1041\substack{+195\\-119}$&5.0&2&$5.2\substack{+0.6\\-0.7}$&0.91&0.05&0.06\\
PSZ2 G108.71-47.75 &$0.389\pm0.001$&&$0.3897\pm0.0002$&11&8&$900\substack{+458\\-190}$&$861\substack{+327\\-216}$&$900\substack{+460\\-183}$&4.3&1&$5.1\substack{+0.7\\-0.8}$&0.99&0.87&0.65\\
PSZ2 G139.62+24.18 &$0.268\pm0.001$&&&20&20&$1120\substack{+366\\-238}$&&$1127\substack{+305\\-171}$&9.6&3&$7.3\pm0.5$&0.51&0.25&0.20\\
PLCK G147.32-16.59 &$0.640\pm0.009$&&&10&10&-- &&-- &5.9&1&$8.1\substack{+0.8\\-0.9}$&0.91&0.91&0.86\\
PSZ2 G157.43+30.34 &$0.402\pm0.001$&+&&28&28&$1244\substack{+192\\-109}$&&$1242\substack{+195\\-103}$&8.8&2&$8.2\pm0.6$&0.99&0.73&0.23\\
PLCK G183.33-36.69 &$0.163\pm0.001$&&&11&11&$897\substack{+437\\-275}$&&$979\substack{+263\\-187}$&2.1&1&$2.3\substack{+0.7\\-0.9}$&0.59&0.05&0.04\\
PSZ2 G186.99+38.65 &$0.377\pm0.001$&&$0.3774\pm0.0003$&41&41&$1506\substack{+164\\-120}$&$1426\substack{+133\\-87}$&$1462\substack{+165\\-102}$&7.1&3&$6.6\substack{+0.6\\-0.7}$&0.83&0.32&0.40\\
PSZ2 G216.62+47.00 &$0.385\pm0.001$&&$0.3864\pm0.0003$&37&37&$1546\substack{+174\\-132}$&$1779\substack{+207\\-153}$&$1524\substack{+178\\-110}$&9.7&3&$8.4\substack{+0.5\\-0.6}$&0.97&0.45&0.86\\
PSZ2 G235.56+23.29 &$0.375\pm0.002$&&&27&23&$1644\substack{+285\\-192}$&&$1636\substack{+294\\-141}$&4.9&3&$5.7\substack{+0.7\\-0.8}$&0.95&0.16&0.13\\
PSZ2 G250.04+24.14 &$0.411\pm0.001$&&&29&29&$1065\substack{+447\\-285}$&&$1466\substack{+380\\-241}$&6.2&3&$6.2\pm0.6$&0.94&0.97&0.10\\
PSZ2 G251.13-78.15 &$0.306\pm0.001$&+&&17&17&$ 801\substack{+852\\-493}$&&$1188\substack{+205\\-155}$&4.8&1&$4.1\pm0.6$&0.56&0.19&0.26\\
PSZ2 G272.85+48.79 &$0.420\pm0.002$&&&10&9&$1462\substack{+389\\-216}$&&$1498\substack{+345\\-175}$&4.8&2&$5.3\substack{+0.7\\-0.8}$&0.98&0.61&0.62\\
PSZ2 G329.48-22.67 &$0.249\pm0.001$&+&&19&16&$ 835\substack{+179\\-119}$&&$746\substack{+152\\-64}$&6.0&3&$5.0\substack{+0.7\\-0.8}$&0.99&0.90&0.46\\
PSZ2 G348.43-25.50 &$0.265\pm0.001$&&&21&20&$1065\substack{+411\\-198}$&&$1160\substack{+277\\-167}$&7.1&3&$6.0\pm0.6$&0.85&0.18&0.02\\
\tableline
PSZ2 G352.05-24.01$^\text{a}$ &$0.786\pm0.026$ &&&23&10&-- &-- &&4.1&1&$6.2\substack{+0.9\\-1.0}$&0.35&0.02&0.03\\
&$0.304\pm0.022$&&&23&13&-- &-- &&&&&0.99&0.94&0.98\\
\tableline \tableline\\
\end{tabular}}
\end{center}
\footnotetext{Two structures observed, not confirmed as clusters (see text and Figure \ref{fig:zhist2}).}
\end{table*}

Masses and $S/N$ were recalculated from a re-extraction of the SZ signal using the Matched Multi-Filter MMF3~\citep{melin2006,PESZ, Planck2014, Planck2015}, fixing the position to the optical position and varying the filter size. They are reported in Table~\ref{specres}. The quoted $S/N$ is the maximum across the various filter sizes at the optical position. The masses are obtained from the re-extracted SZ signal following the method described in Sec. 7.2.2 of~\cite{Planck2014}.

In Table \ref{specres}, we also show the number of detection methods from \citet{Planck2015}.
The Planck selection function  is very reliable ($>90\%$) for detections obtained with $S/N>4.5$ by at least one detection method.  For objects detected with all three detection methods, the probability of being a cluster is $>98\%$ with $S/N>4.5$  \citep{Planck2015}. In order to confirm each target as galaxy cluster, we combine this information with the probability that the galaxy redshift distribution is Gaussian, the characteristic distribution of a virialized cluster, from the Kolmogorov-Smirnov \citep[K-S, e.g.][]{ff1987} and the Shapiro-Wilk \citep[S-W,][]{sw1965} statistics, as well as the probability of a uniform distribution from a K-S test. The results of these tests are shown in the last three columns of Table \ref{specres}.

Eleven of our cluster candidates have $> 98\%$ probability of being a galaxy cluster, since they were detected with three detection methods and have $S/N > 4.5$. For these targets, the probabilities that the redshift distributions are Gaussian are almost always $> 80\%$ and the probabilities to be uniform always $<50\%$ and mostly $<10\%$. Only one object, PSZ2 G139.62+24.18 at z=0.268, has a $S/N=9.5$, which corresponds to a \Planck\ reliability of being a cluster of $\sim100\%$, but a K-S (S-W) probability of having a Gaussian redshift distribution of $\sim 50\%$ ($\sim 20\%$), and the probability of having a uniform redshift distribution of $\sim 20\%$. It shows a very luminous BCG at the center, and has 20 spectroscopically confirmed galaxies at the same redshift. All these elements lead us to believe that this is a galaxy cluster, and it was also confirmed as a cluster in the PSZ2 catalog. All the other ten targets are mostly likely galaxy clusters, and we assume that they are. Of those, we confirm three clusters that were not originally confirmed in the PSZ2.

The other cluster candidates that were detected with at least one detection method and $S/N > 4.5$ have a $>90\%$ probability of being galaxy clusters. For these candidates, we assume that we confirm a cluster when the probability that their redshift distribution is Gaussian is $>95\%$ ($\sim 2\sigma$). On the other hand, we do not confirm a cluster when the probability of a uniform distribution is $>50\%$. In fact, since the \Planck\ detection and the galaxy redshift distribution are two independent events, we can multiply the \Planck\ probability of not being a cluster ($\sim 10\%$) by the probability of having a uniform distribution of galaxy redshifts. If this last is $<50\%$, the total probability that the candidate is not a cluster is $<5\%$. 
Among these last targets, three have a  probability that their redshift distribution is  Gaussian is $>95\%$ ($\sim 2\sigma$), and we consider them as confirmed clusters. All three are new confirmation with respect to PSZ2. 

Three of the targets that were only detected by one method, though, and one candidate detected with two methods show less definitive results. We discuss these last cluster candidates in more detail below.

PLCK G147.32-16.59 was detected by one method with a high $S/N$ ($S/N \sim 6$), and its redshift distribution has a probability of $\sim 90\%$ of being Gaussian; however, it also has a $\sim 10\%$ probability of not being a cluster. With only 10 confirmed members, its confirmation is not very reliable, but it is more probable that it is a cluster or a group of galaxies than a uniform redshift distribution, and we consider it a confirmed cluster.  \XMM\ observations \citep{planck_int_iv} reveal two substructures in the X-ray surface brightness, indicating that it is undergoing a merger event \citep[see also][]{vanweeren+14,mroczkowski+15}. Because of the undergoing merger, we have excluded this cluster from the analysis of the velocity dispersion--mass relation in \citet{amodeo+17}.

PLCK G183.33-36.69 was detected by one method with a $S/N \sim 2$ (\Planck\ reliability of $<70\%$), its redshift distribution has a K-S (S-W ) probability of $\sim 60\%$ ($\sim 5\%$) to be Gaussian, and a $\sim 1\%$ total probability of not being a cluster. However, we can clearly see the two bright central galaxies in the Gemini image, and the cluster center is close to the border of the Gemini field. It seems to us that this cluster was not enough well centered in the Gemini imaging and spectroscopy to obtain a significant sample to confirm it, even if it has a larger probability to be a cluster or group of galaxies instead of an uniform galaxy distribution. The SZ flux gives a mass of $\mMpl = 2.3\substack{+0.7\\-0.9}  \times 10^{14} M_{\odot}$, and its galaxy velocity dispersion is $\sigma_{200} = 842\substack{+297\\-451}$ km~$s^{-1}$. 
We consider it as a confirmed cluster, and warn the reader about the larger uncertainty (with respect to most of the remaining sample) in the velocity dispersion measurement and its redshift distribution skewness, which both might indicate an unrelaxed dynamical state. We kept this cluster in our sample in \citet{amodeo+17} because, due to the large uncertainty on the velocity dispersion measurement, it does not significantly weight on our final results.

PSZ2 G251.13-78.15 was detected by one method with a $S/N \sim 4.8$ (\Planck\ reliability of $\sim90\%$), its redshift distribution has a K-S and a S-W probability of $\sim 60\%$ and $\sim 20\%$, respectively, to be Gaussian, and a $\sim 3\%$ probability of not being a cluster. We consider it as a confirmed cluster, and again notice the larger uncertainty in its confirmation, mass and velocity dispersion estimates. This is a newly spectroscopically confirmed cluster.

PSZ2 G272.85+48.79 was detected by two methods with a $S/N \sim 5$ (\Planck\ reliability of $\sim92\%$). From the combined Planck and K-S gaussian probabilities, it has a ~90\% probability of being a cluster. On the other hand, from the combined Planck and K-S uniform probabilities, it has a ~5\% of probability of not being a cluster. According to our criteria this is at the limit of being confirmed as a cluster of galaxies. However, we assume it is confirmed, also considering that it is more massive than $10^{14} M_\odot$ \citep[e.g.][]{evrard+2008}. 

For PSZ2 G352.05-24.01, the redshift obtained from the X-ray analysis is $z=0.79$ \citep{planck_int_iv}, but we observe galaxies in a wider redshift range. In fact, we can distinguish two structures at $z\sim0.8$ and $z\sim0.3$,  shown in blue and green, respectively, in Figure \ref{fig:zhist2}.  Both redshift distributions have a standard deviation of $\sim 0.08$, much wider of what expected for a cluster of galaxies.  This target is not a cluster of galaxies, and we have excluded it from the analysis of the velocity dispersion--mass relation in \citet{amodeo+17}. 

PSZ2 G085.95+25.23, confirmed at $z=0.782\pm0.010$, is one of the highest redshift confirmed \Planck\ clusters.

Newly confirmed clusters are labeled with the sign "+" in Table \ref{specres}.

\section{Spectroscopic redshift catalogs}
\label{sec:catalog}

We provide the cluster catalogs as electronic documents, including the following parameters for each cluster galaxy:
\begin{enumerate}
\item the galaxy identification number ID
\item the J2000 right ascension R.A., in hours
\item the J2000 declination Decl., in deg
\item the measured spectroscopic redshift SPECZ
\item the error in spectroscopic redshift eSPECZ
\end{enumerate}
An example is shown in Table \ref{tab:catalog} for PSZ2G053.44-36.25. 

\section{Discussion}
\label{sec:discussion}

In the context of the optical identification of \Planck\ cluster candidates, our sample, although small, is chosen to have a wide range of mass with the aim of obtaining a statistical calibration of the \Planck\ SZ mass estimator. In this section, we compare it with previous \Planck\ cluster redshift measurements.

Eight of our targets are in the SDSS DR8 redMaPPer cluster catalogs \citep{sdss3,rykoff+14}. Five of them (PSZ2G108.71-47.75, PSZ2 G186.99+38.65, PSZ2 G216.62+47.00, PSZ2 G056.93-55.08, and PSZ2 G083.29-31.03) have previous spectroscopic redshift measurements in agreement with our values.

Measurements of galaxy redshifts are available in the SDSS DR14 in seven of our fields. A search within two virial radii from the center of each of our clusters finds spectroscopic catalogs for galaxies in the following clusters ($\text{N}_\text{gal, DR14}$): 
PSZ2 G053.44-36.25 (15),
PSZ2 G056.93-55.08 (3),
PSZ2 G081.00-50.93 (8),
PSZ2 G083.29-31.03 (32),
PSZ2 G108.71-47.75 (3),
PSZ2 G186.99+38.65 (24),
PSZ2 G216.62+47.00 (19).
We included these redshifts and we re-calculated the cluster redshifts and velocity dispersions with the same procedure (see Table~\ref{specres}). The redshift estimates do not change, while the uncertainties are smaller. Velocity dispersions are on average within ($0.28\pm 0.17$) sigma the values obtained with our measurements only, and never above 0.5 sigma.
For the other targets, there are not public spectroscopic redshifts for single galaxies to our knowledge.

The \Planck\ collaboration has undertaken two important optical follow-up programs to confirm \Planck\ cluster candidates and to measure their redshifts. The first is based on observations with the Russian-Turkish 1.5 m telescope \citep{planck_int_xxvi} and provides spectroscopic redshifts for 65 \Planck\ clusters. It includes our targets PSZ2 G139.62+24.18, for which they obtain a spectroscopic redshift of 0.268 consistent with our measurement, and PSZ2 G157.43+30.34, for which they find a photometric redshift of 0.45.  \citet{vorobyev+16} report on additional spectroscopic observations of the latter cluster from the 2.2-m Calar Alto Observatory telescope, obtaining $z=0.403$ with an error of 1\%, consistent with our value of $z=0.402\pm0.006$.

The second program, based on observations with telescopes at the Canary Islands Observatories (Gran Telescopio Canarias, Isaac Newton Telescope, William Herschel Telescope, Telescopio Nazionale Galileo, Nordic Optical Telescope, IAC80 telescope), provided 53 cluster spectroscopic redshifts, and is published in \citet{planck_int_xxxvi}.
Again it includes our target PSZ2 G139.62+24.18, for which they measure z=0.266 from 22 spectroscopically confirmed members, consistent with our value of $z=0.268\pm0.005$ obtained from 20 galaxies.

The \Planck\ collaboration has also carried out X-ray validation programs with \XMM\ \citep{planck_early_ix,planck_int_i,planck_int_iv}, where redshifts $z_\text{Fe}$ have been obtained from X-ray spectral fitting of the iron emission line. Targets PSZ2 G250.04+24.14 and PSZ2 G272.85+48.79 are analysed in \citet{planck_early_ix}, PSZ2 G235.56+23.29 in \citet{planck_int_i}, and PSZ2 G348.43-25.50 and PLCK G147.32-16.59 in \citet{planck_int_iv}. In all cases, \XMM\ finds  redshifts consistent with our values. 
\citet{planck_int_iv} also includes the X-ray analysis of PSZ2 G329.48-22.67. They observe a double projected system at redshifts 0.24 and 0.46. In our GMOS analysis, we measure $z=0.249\pm0.003$ based on 16 spectroscopic members, with no detections at higher redshift. 

Finally, \citet{planck_int_iv} quote a redshift $z_\text{Fe}=0.77$ for PSZ2 G352.05-24.01. The authors give $z_\text{Fe}=0.12,\,0.40$ as other possible solutions to the spectral fitting, but these are excluded from the comparison between the X-ray and SZ  properties of the source ($Y_\text{X}/Y_\text{500}$). We observe a sparse galaxy distribution, with two (small) peaks with more than five galaxies, one with six galaxies at $z=0.798\pm0.021$ and the other with 11 at $z=0.334 \pm 0.025$. However, these large dispersions ($\sim$ 3500~km/s at z=0.798 and $\sim$ 5600~km/s at z=0.334) do not confirm clusters of galaxies, and we do not consider this target as a confirmed cluster.

In conclusion, six of our clusters have spectroscopic redshifts from previous optical studies, seven have redshift measurements from X-ray spectral fitting. Their velocity dispersions are published in this paper for the first time. For the remaining seven clusters, spectroscopic redshifts and velocity dispersions are published in this paper for the first time.

\section{Conclusions}
\label{sec:conclusions}

This article presents spectroscopic redshifts and velocity dispersions for 20 \Planck\ SZ clusters. We spectroscopically confirm 19 clusters with Gemini-North and Gemini-South/GMOS, six of which were spectroscopically confirmed in this paper for the first time. 
We also confirm and measure the redshift and velocity dispersion of the \Planck\ cluster PSZ2 G085.95+25.23  with Keck/LRIS spectroscopy, measuring a mean redshift of $z=0.782\pm0.010$, one of the \Planck's  highest redshift confirmed clusters.
Eighteen of our clusters are included in the last released \Planck\ SZ source catalog, PSZ2 \citep{Planck2015}.

We provide online catalogs for each cluster spectroscopic member redshift (an example is shown in Table \ref{tab:catalog}).

In a companion paper \citep{amodeo+17} we use the cluster galaxy velocity dispersions to measure the \Planck\ mass bias, and to constrain the cluster velocity bias.

\acknowledgments{Based on observations obtained at the Gemini Observatory (Programs GN-2011A-Q-119, GN-2011B-Q-41, and GS-2012A-Q-77; P.I. J.G. Bartlett), which is operated by the Association of Universities for Research in Astronomy, Inc., under a cooperative agreement with the NSF on behalf of the Gemini partnership: the National Science Foundation (United States), the National Research Council (Canada), CONICYT (Chile), Ministerio de Ciencia, Tecnolog'a e Innovaci—n Productiva (Argentina), and MinistŽrio da Cincia, Tecnologia e Inova‹o (Brazil). Supported by the Gemini Observatory, which is operated by the Association of Universities for Research in Astronomy, Inc., on behalf of the international Gemini partnership of Argentina, Brazil, Canada, Chile, and the United States of America. 
This material is based upon work supported by AURA through the National Science Foundation under AURA Cooperative Agreement AST 0132798 as amended. We are pleased to acknowledge the Palomar Observatory staff for their enthusiastic and excellent support. 
Part of the data presented herein were obtained at the W. M. Keck Observatory, which is operated as a scientific partnership among the California Institute of Technology, the University of California and the National Aeronautics and Space Administration. The Observatory was made possible by the generous financial support of the W. M. Keck Foundation. The authors wish to recognize and acknowledge the very significant cultural role and reverence that the summit of Maunakea has always had within the indigenous Hawaiian community.  We are most fortunate to have the opportunity to conduct observations from this mountain. We thank the P.I. of the Keck observations, Fiona Harrison, and Mislav Balokovi\'c and George Lansbury for participating in the Keck observations.
J.G.B. and S.M. acknowledge financial support from the {\em Institut Universitaire de France (IUF)} as senior members. 
Part of the work of J.G.B., C.L. and D.S. was carried out at the Jet Propulsion Laboratory, California Institute of Technology, under a contract with NASA. S.M.'s research was supported by an appointment to the NASA Postdoctoral Program at the Jet Propulsion Laboratory, administered by Universities Space Research Association under contract with NASA. We thank the referee for her/his useful comments that helped improve the presentation of this work, and Joanne Cohn for useful discussion.}

\facility{Gemini: South, Gemini: Gillet, Hale, Keck:I (LRIS), Planck}
\newpage
\bibliographystyle{aasjournal}
\bibliography{mybib}

\appendix  
\section{\Planck\ clusters observed with the Palomar telescope} \label{sec:palomar}
\begin{table*}[h!]
\begin{center}
\caption{\Planck\ clusters observed with the Palomar telescope. Clusters are named after their PSZ2 or PSZ1 ID, when available. When it is not available, we use the prefix 'PLCK' followed by a notation in Galactic coordinates similar to that used in the PSZ2 paper.  }
\label{tab:palomar}
\vspace{0.25cm}
\resizebox{!}{11cm}{
\begin{tabular}{llccccccccccccccc}
\tableline \tableline\\
Name&R.A.& Decl.&Filter&Instrument & Run\\
&(deg)&(deg)&\\
\tableline \tableline 
PSZ2 G019.12+31.23 & 249.1420 & 3.1528 & \it g', i' & LFC & 2011A \\
PLCK G024.20+58.78 & 225.5920 &18.6586 & \it g', i' & LFC & 2011A \\
PLCK G030.89+42.25 & 243.3310 & 16.4481 & \it g', i', r' & LFC & 2011A \\
PSZ2 G066.41+27.03 & 269.2120 & 40.1156 & \it g', i', r' & LFC & 2011A \\
PLCK G071.59-63.16 & 351.9458 &	 -8.9647 &\it   i' & LFC & 2012B \\
PSZ2 G074.08-54.68 & 347.0917 &	-1.9106 &\it   i' & LFC & 2012B  \\
PSZ2 G078.67+20.06 & 282.9920 & 49.0257 & \it g', i' & LFC & 2011A \\
PSZ2 G082.31-67.00 & 357.9500 & -8.9647&\it i' & LFC & 2012B\\
PSZ2 G086.93+53.18 & 228.4790 & 52.7775 & \it g', i' & LFC & 2011A \\
PLCK G087.67+23.00 & 282.3250 & 57.8956 & \it g', i' & LFC & 2011A \\
PSZ2 G091.83+26.11 & 277.8080 & 62.2317 & \it i', r' & LFC & 2011A \\ 
PSZ2 G094.56+51.03 & 227.0960 & 57.8706 & \it g', i', r' & LFC & 2011A \\
PLCK G096.88+24.22 & 284.0750 & 66.3819 &\it g', i' & LFC & 2011A \\
PSZ2 G107.83-45.45 & 1.8753  & 16.1423 & \it g', r', J, K & LFC, WIRC & 2010B \\
PSZ1 G108.52+32.30 & 256.9920 & 76.4697 & \it g', r', i' & LFC & 2011A \\
PLCK G109.35+64.36 & 202.3080 & 51.7589 & \it g', i' & LFC & 2011A \\
PLCK G113.07-74.37 & 10.1610 & -11.7062 & \it r', J, K & LFC, WIRC & 2010B \\
PLCK G113.66+70.59 & 197.2970 & 46.2171 & \it g', i' & LFC & 2011A \\
PLCK G114.92-20.06 & 2.6792  &	42.1783 &\it   i'  & LFC & 2012B \\ 
PLCK G116.80-25.18 & 5.8708  &	37.3600 &\it   i'  & LFC & 2012B\\
PLCK G117.14-26.47 & 6.4417 & 36.1117 &\it   i'  & LFC & 2012B \\
PSZ1 G121.09+57.02 & 194.8400 & 60.0897 &\it g', i', r' & LFC & 2011A \\
PSZ1 G129.07-24.12 & 20.0000 & 38.4531	&\it   g', i'  & LFC & 2011B \\
PSZ2 G134.26-44.28 & 21.3542 &	17.8808 &\it   g', i'  & LFC & 2011B \\
PSZ2 G138.11+42.06  &157.0542 &	70.6081 &\it g', i', r', J, K & LFC, WIRC & 2010B, 2011A, 2011B \\
PSZ2 G139.62+24.18 & 95.4912 & 74.7042 & \it g', i', r', J, K & LFC, WIRC &2010B, 2011A \\
PLCK G142.35+17.59 & 78.8752 & 69.7009 & \it g', J & LFC, WIRC & 2010B, 2011A \\
PLCK G147.32-16.59$^\text{a}$  & 44.1000 & 40.2911 &\it   g', i', r' & LFC & 2011B \\
PSZ2 G157.43+30.34 & 117.2208 &	59.6944 &\it   g', i', r', J, K & LFC, WIRC & 2010B, 2011A, 2011B \\
PLCK G159.41-62.64 &28.7625 &	-4.3600 &\it   g'  & LFC & 2011B\\
PSZ2 G171.98-40.66 & 48.2307 & 8.3805 & \it g', r', K & LFC, WIRC & 2010B\\
PSZ2 G172.93+21.34 & 106.8920 & 44.3050 & \it r' & LFC & 2011A \\
PLCK G183.33-36.69 & 57.2936 & 4.5974 & \it g', J, K & LFC, WIRC & 2010B \\
PSZ2 G183.30+34.98 & 127.4042 &	38.4325 &\it   g', i' & LFC &  2011B\\
PLCK G184.34+29.07 & 120.3380 & 36.4269 & \it g', i' & LFC & 2011A \\
PSZ2 G193.31-46.13 & 53.9592 & -6.9853 & \it g', r', J, K & LFC, WIRC &  2010B \\
PSZ2 G193.63+54.85  &152.5750 &	32.8472 &\it   i'  & LFC & 2011B\\
PSZ2 G194.68-49.76 & 51.3625 & -9.6181 &\it   i' & LFC &  2012B\\
PSZ2 G196.65-45.51 &  55.7583 & -8.7039 &\it   i'  & LFC & 2012B\\
PLCK G196.72+23.27 & 118.2330 & 24.2689 & \it g', i' & LFC & 2011A \\
PLCK G198.13-24.68 & 74.3315 & 0.9310 & \it r', J, K & LFC, WIRC & 2010B \\
PSZ2 G198.90+18.16 & 113.4333 &	20.3083 &\it  g', i'  & LFC  &2011B\\
PLCK G201.89+32.14 & 128.5292 & 22.7656 &\it   g', i'  & LFC & 2011B\\
PSZ1 G203.88+62.50 & 161.7580  & 27.9606 & \it i' & LFC & 2011A \\
PSZ2 G204.24+14.51 &   112.1375 & 14.1283 &\it  g', i'  & LFC & 2012B\\
PSZ2 G205.90+73.76 & 174.5833 &	27.9186 &\it   g', i'   & LFC & 2011B \\
PLCK G211.37+49.36 & 148.5583 & 21.2128 &\it   g', i' & LFC &2011B \\
PSZ2 G212.44+63.19 & 163.2292 &	24.2128 &\it g', i'  & LFC & 2011B\\
PLCK G214.57+36.96 & 137.1950 & 14.7084 &\it r' & LFC & 2011A \\
PLCK G219.13+52.94 & 153.8580 & 17.8178 & \it g', i', r', J, K & LFC, WIRC & 2010B, 2011A \\
PSZ1 G223.80+58.50 & 160.3292 &	17.5111 &\it   g', i'  & LFC & 2011B\\
PLCK G247.33+63.53 & 170.8870 &10.6117 & \it g', i' & LFC & 2011A \\
PSZ1 G263.75+53.85 & 170.9875 &	-2.2161 &\it   g', i'  & LFC & 2011B\\
\tableline \tableline \\
\end{tabular}}
\footnotetext{Target PLCK G147.32-16.59 is confirmed in the \XMM\ cluster validation \citep{planck_int_iv}, but it is not included in the two \Planck\ catalogs of SZ sources released so far.}
\end{center}
\end{table*}

\newpage

\section{Redshift histograms, optical images and SZ maps}
\label{sec:figures}

\begin{figure*}[h!]
\begin{center}
\figurenum{B1}
\epsscale{1.2}
\plotone{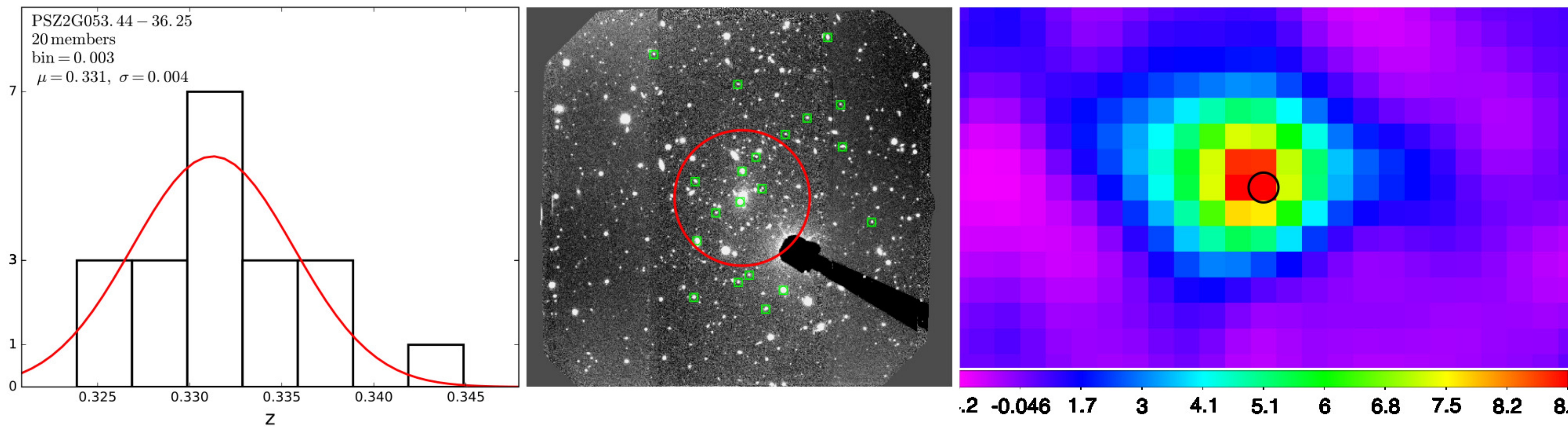}
\plotone{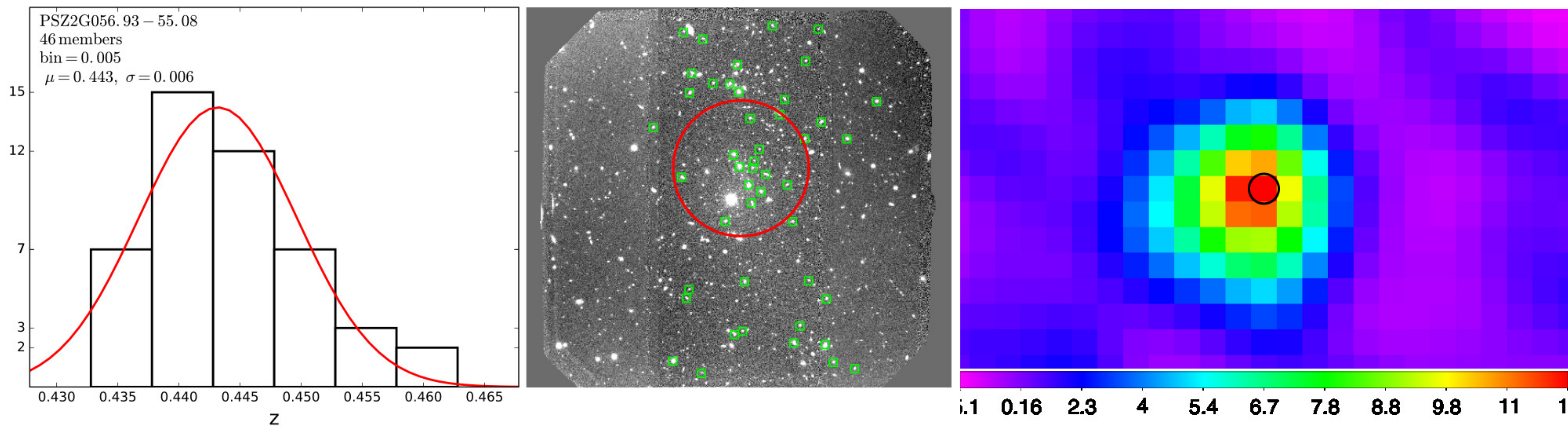}
\plotone{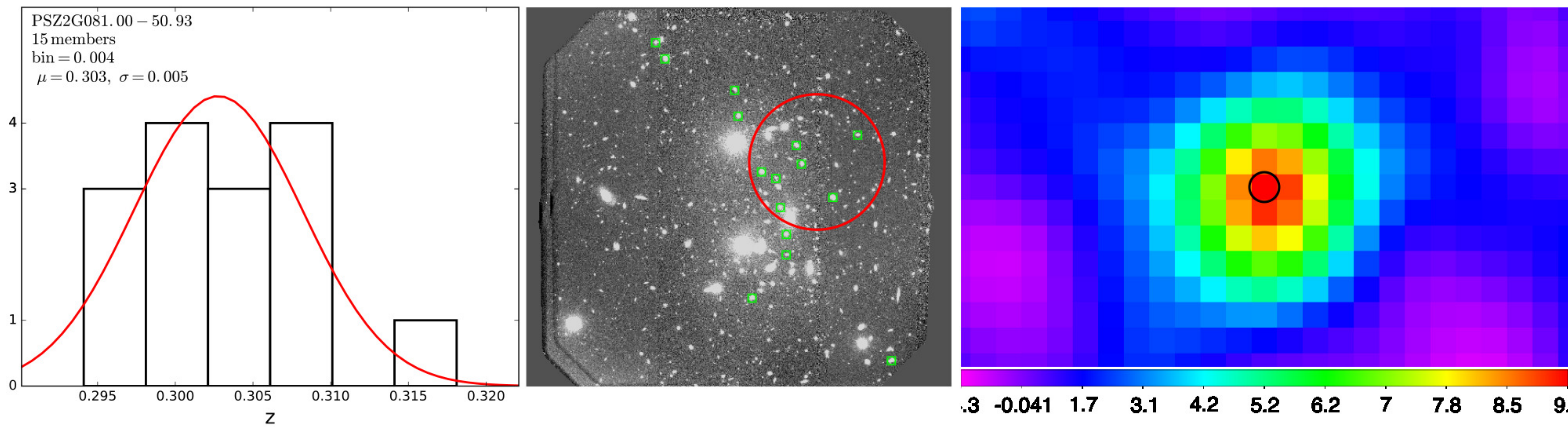}
\plotone{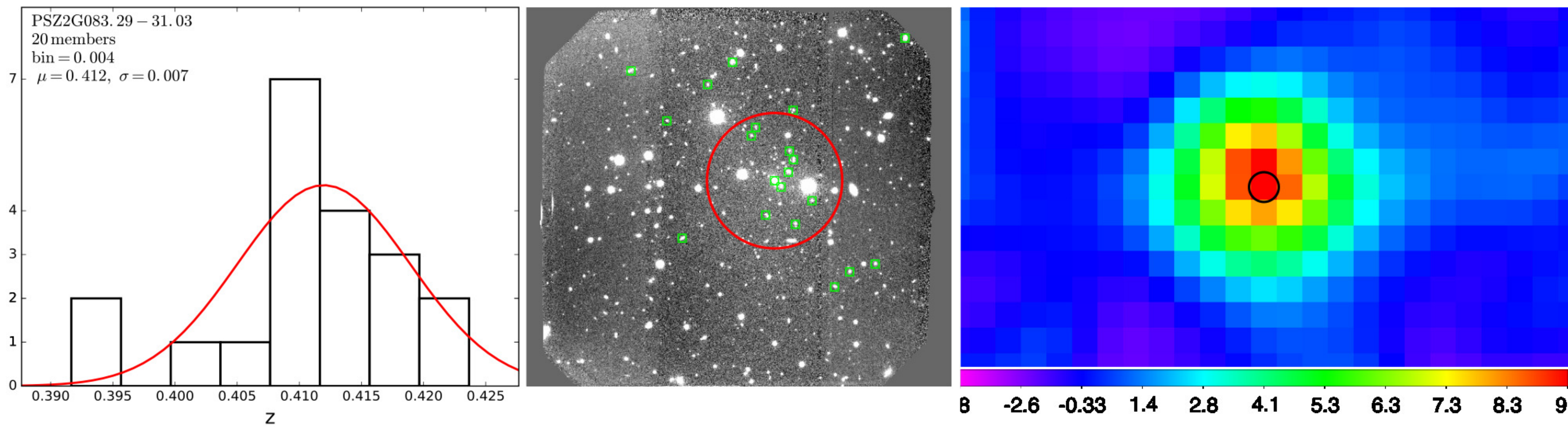}
\end{center}
\text{\it Figure B1 continued}
\end{figure*}
\clearpage

\begin{figure*}[h!]
\begin{center}
\figurenum{B1}
\epsscale{1.2}
\plotone{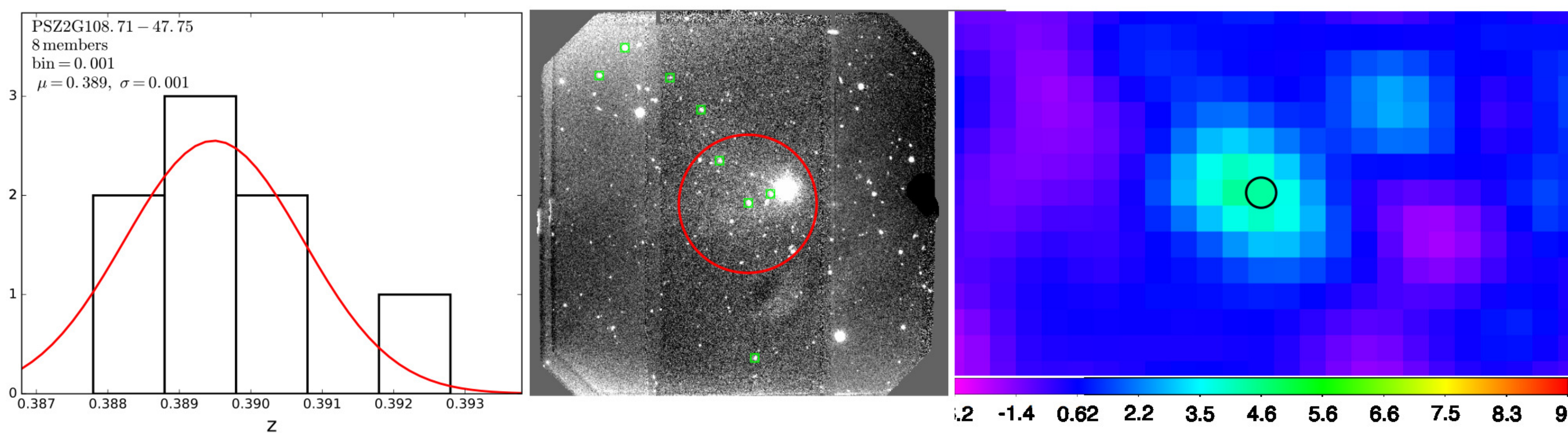}
\plotone{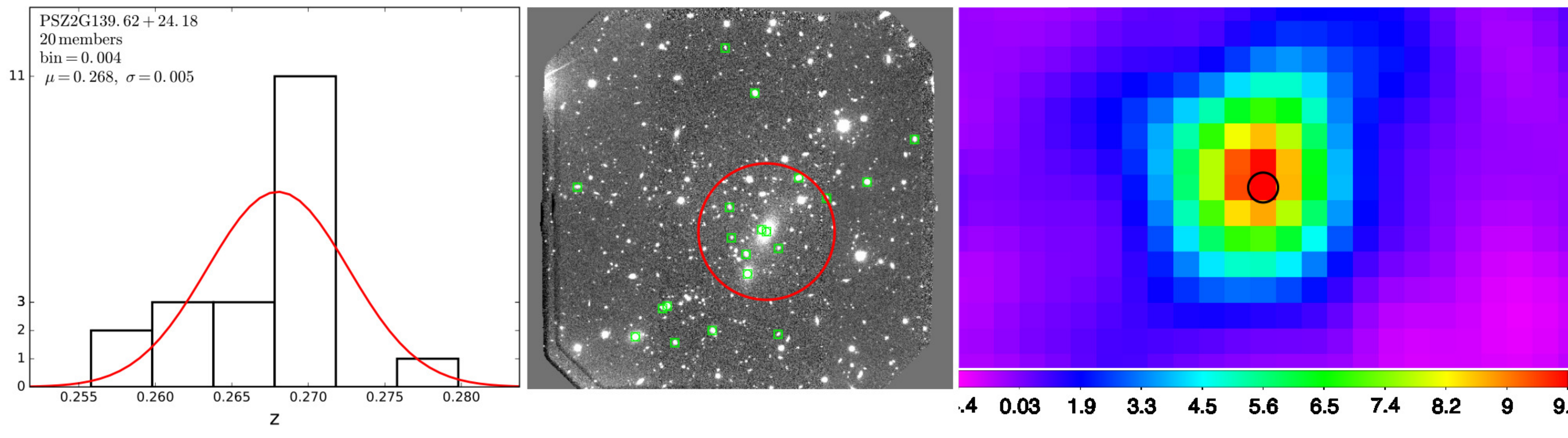}
\plotone{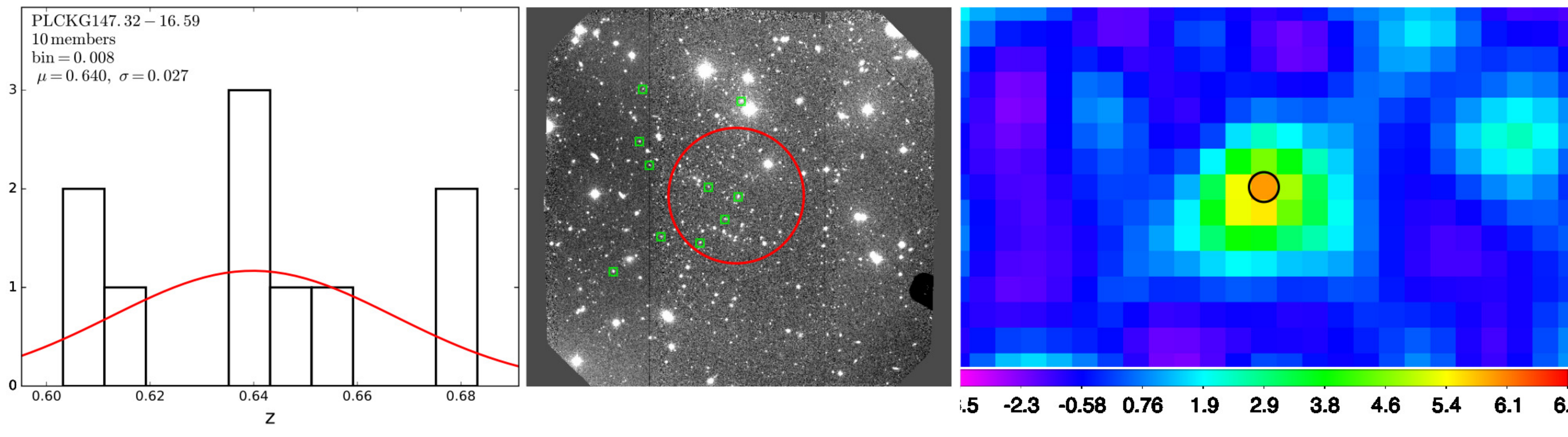}
\plotone{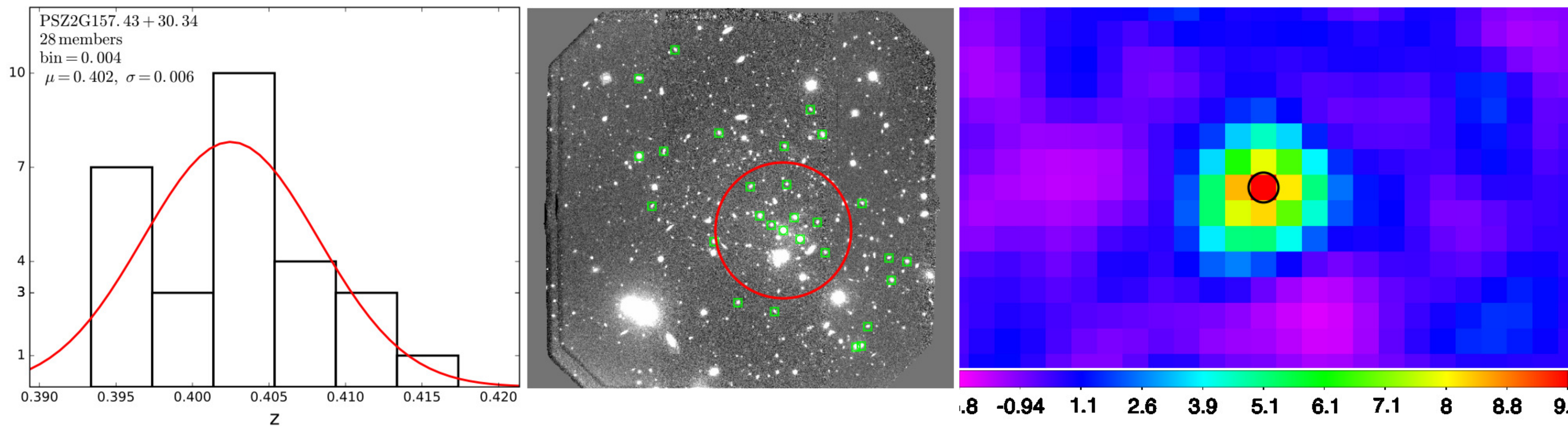}
\end{center}
\text{\it Figure B1 continued}
\end{figure*}
\clearpage

\begin{figure*}[h!]
\begin{center}
\figurenum{B1}
\epsscale{1.2}
\plotone{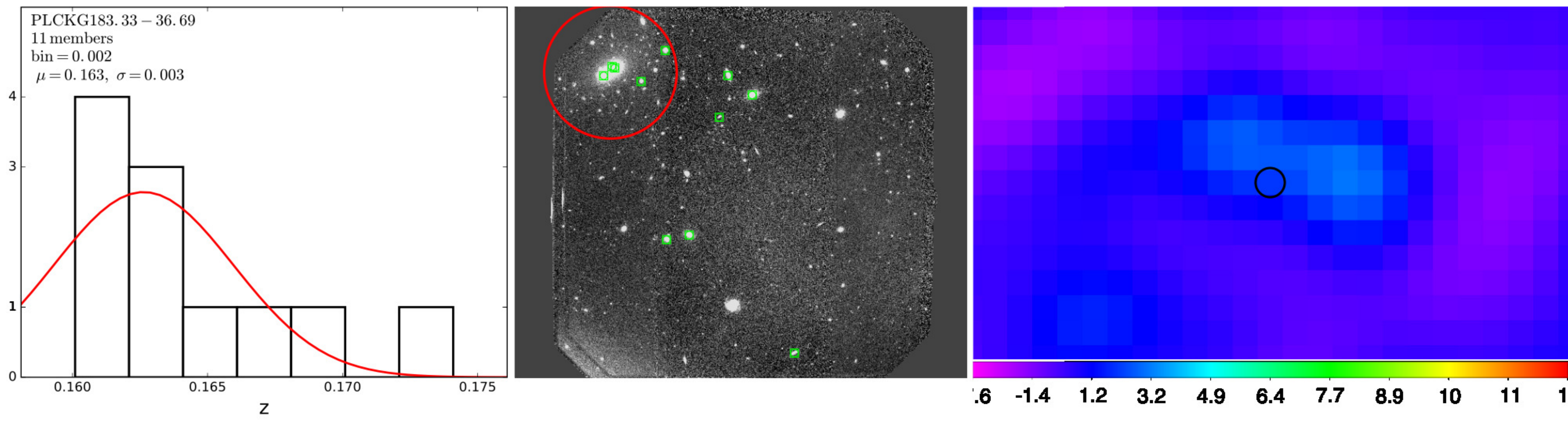}
\plotone{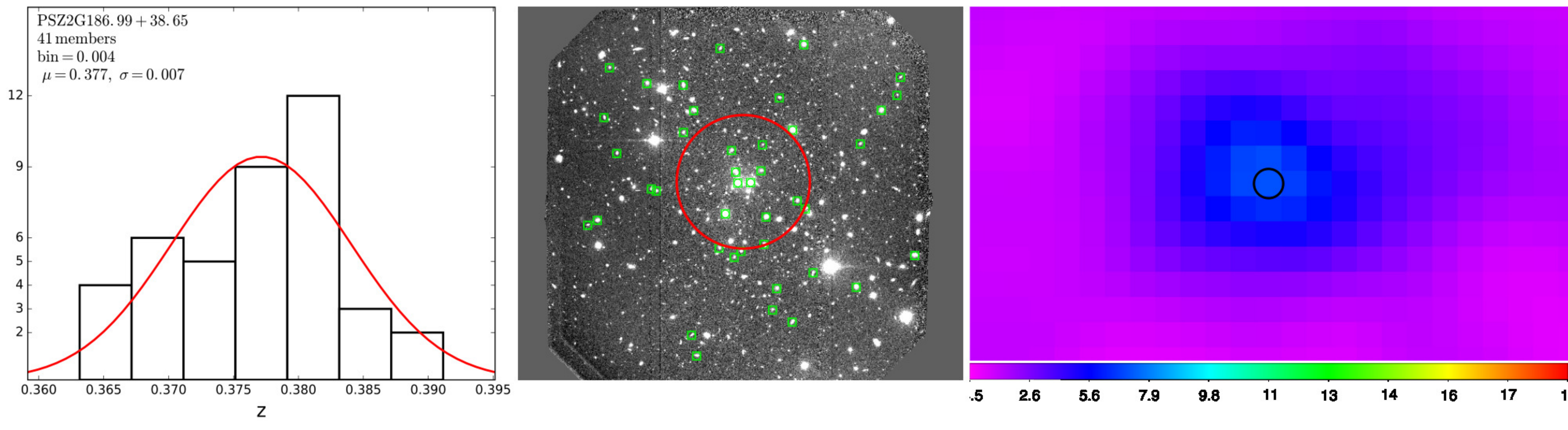}
\plotone{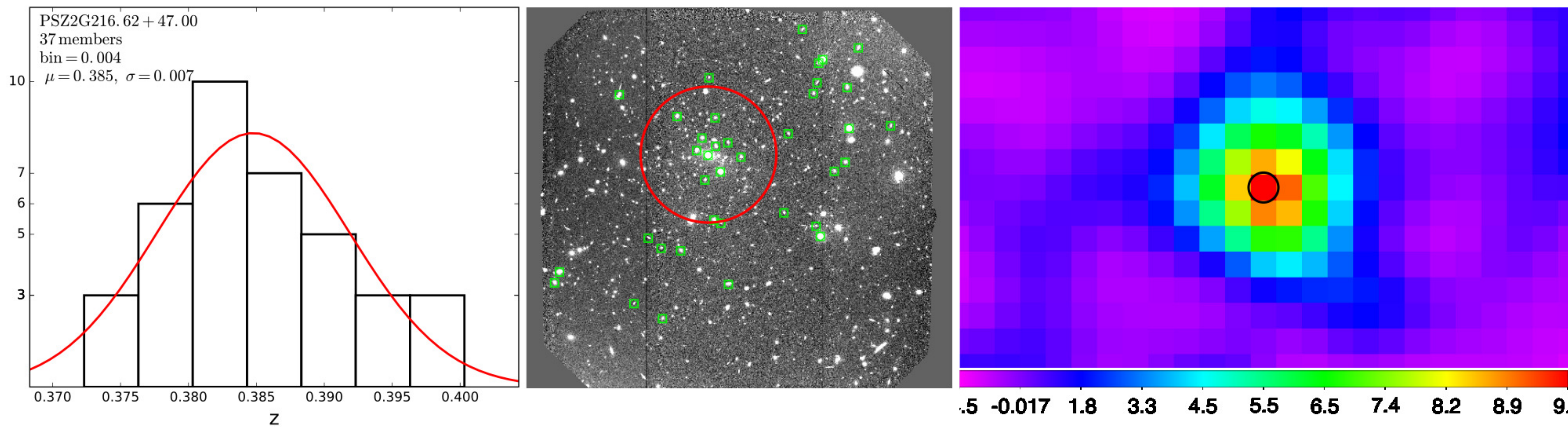}
\caption{Redshift histograms (left), optical images (middle) and SZ maps in signal-to-noise units (right) of clusters in the Northern sample. The red curve in the histograms is a Gaussian fit with mean ($\mu$) and standard deviation ($\sigma$) indicated in the legends, calculated for the redshift distribution using the biweight method. We also indicate the number of members in each cluster and the size of the redshift bins. The red (black) circles in the images encloses a circle of radius 1 arcmin around the optical (SZ) center of the clusters, while the confirmed member galaxies are shown by green squares.}
\label{fig:zhist}
\end{center}
\end{figure*}

\begin{figure*}[h!]
\begin{center}
\figurenum{B2}
\epsscale{1.2}
\plotone{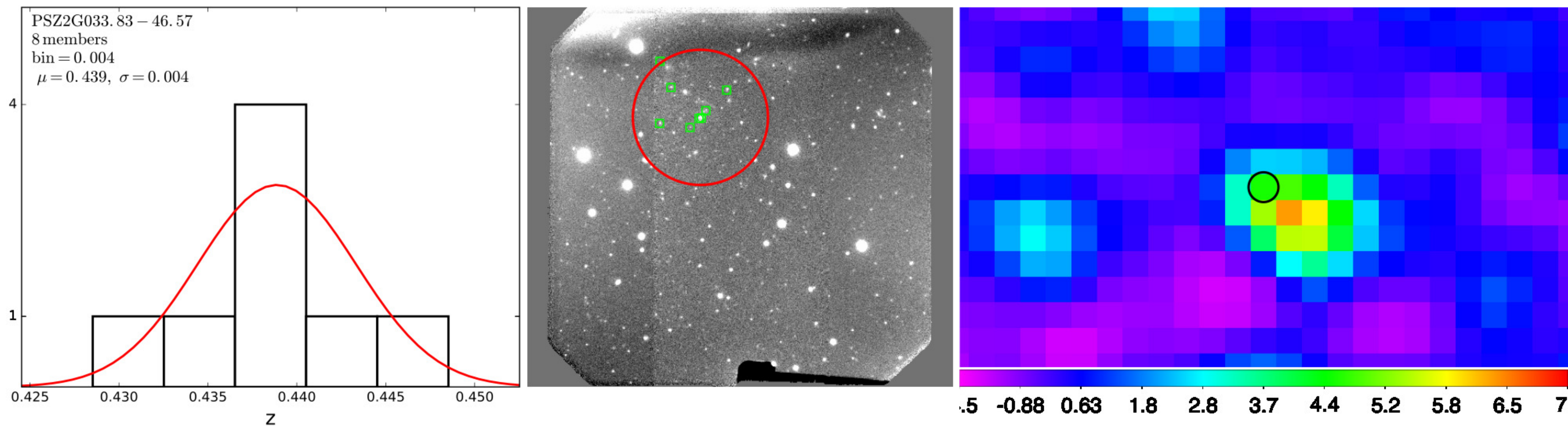}
\plotone{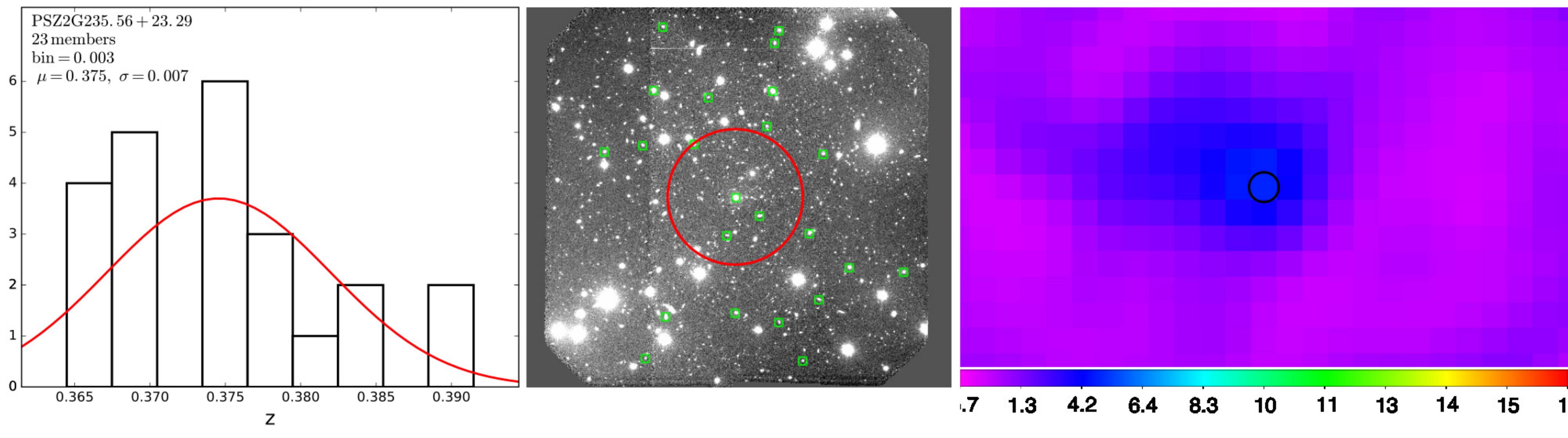}
\plotone{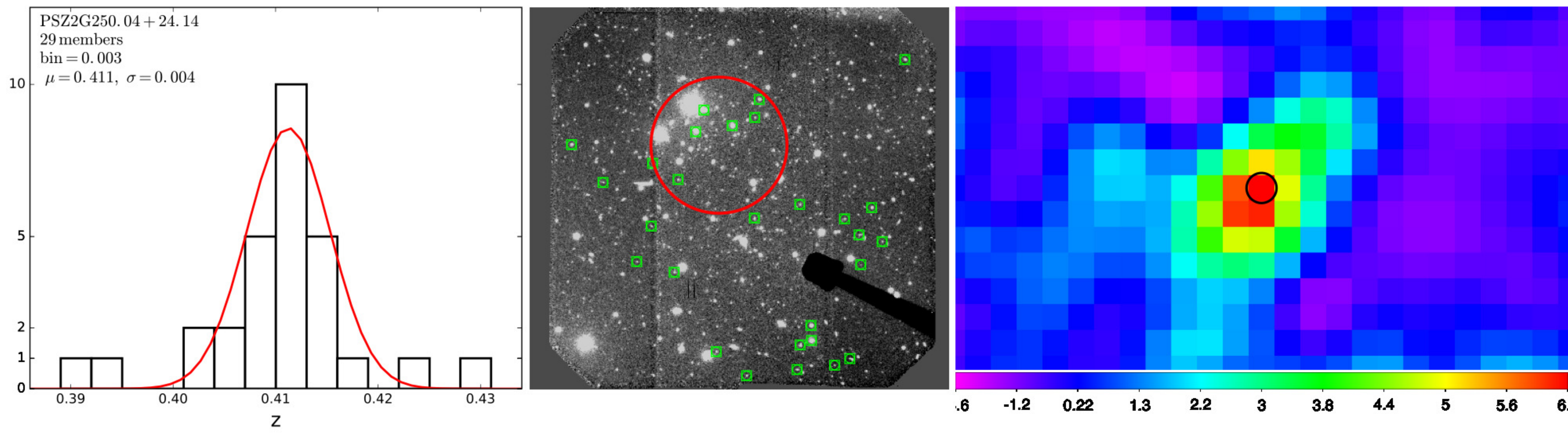}
\plotone{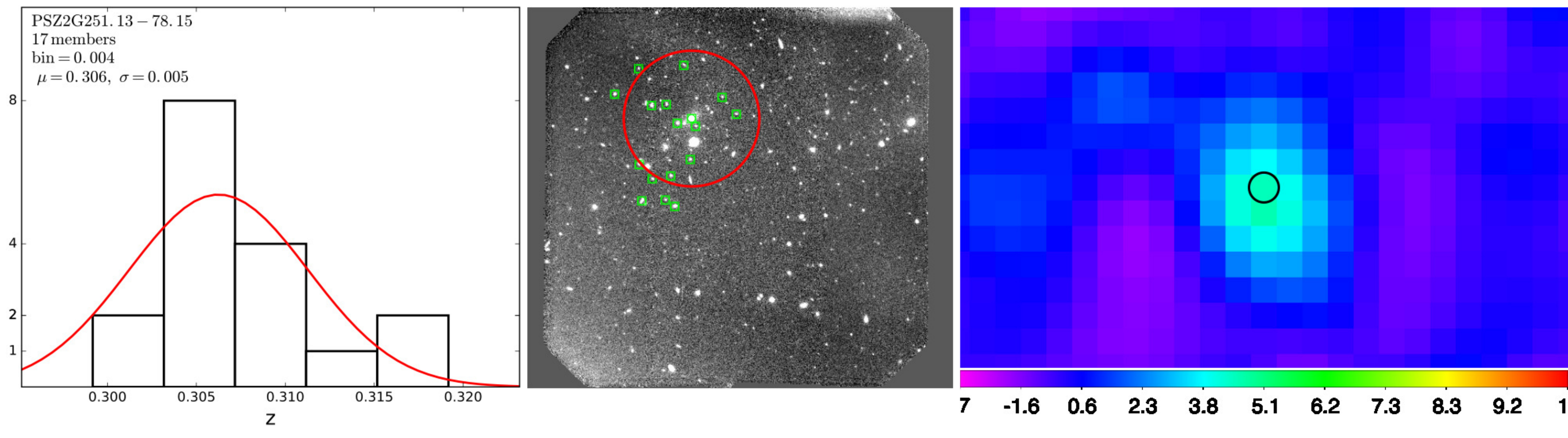}
\end{center}
\text{\it Figure B2 continued}
\end{figure*}
\clearpage

\begin{figure*}[h!]
\begin{center}
\figurenum{B2}
\epsscale{1.2}
\plotone{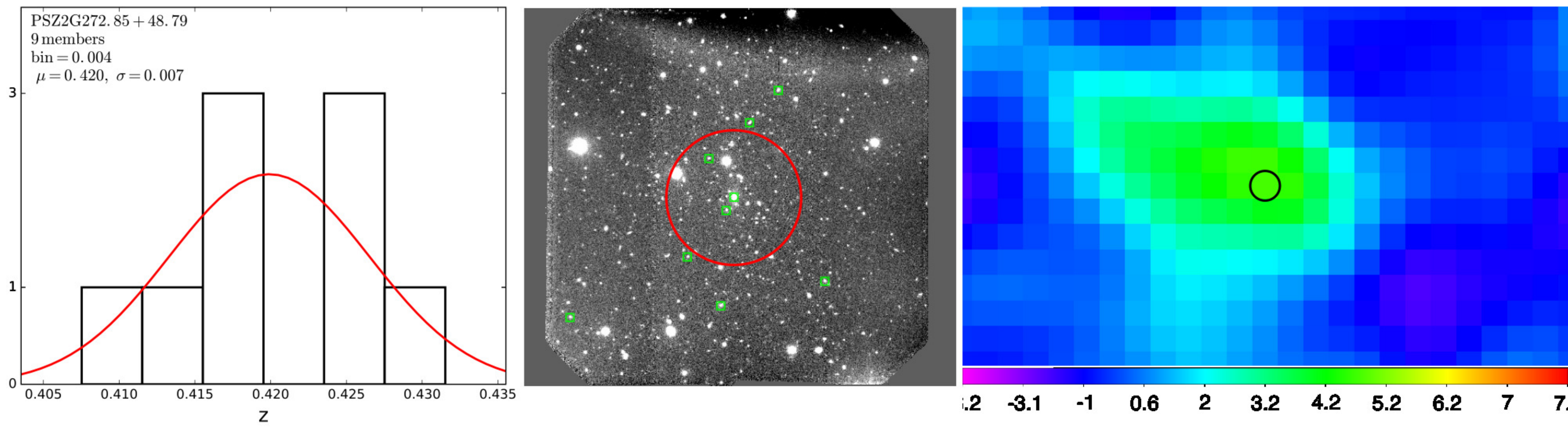}
\plotone{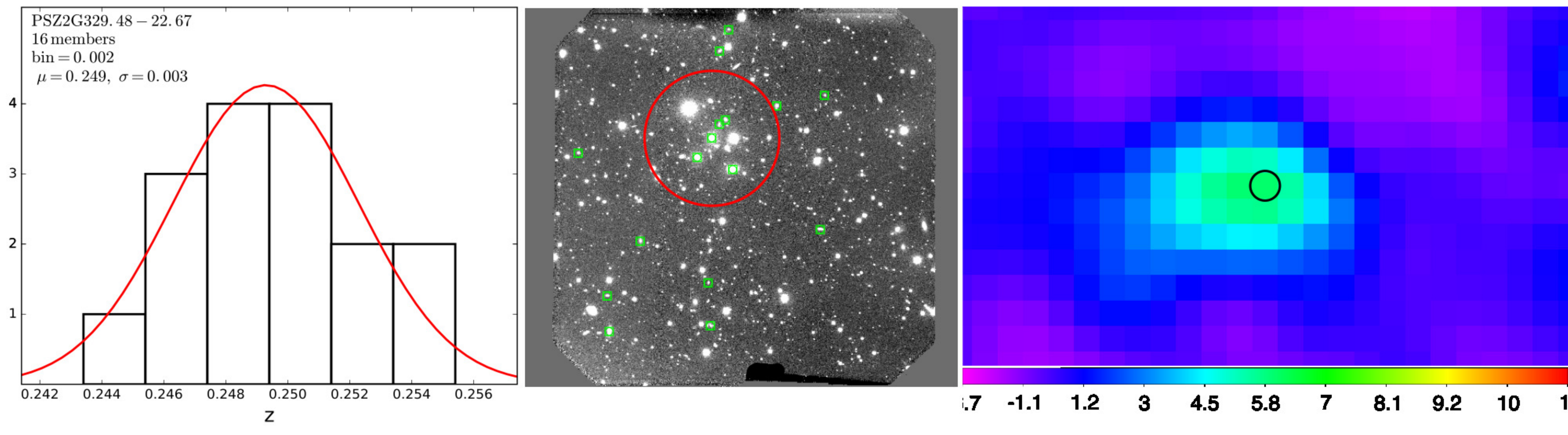}
\plotone{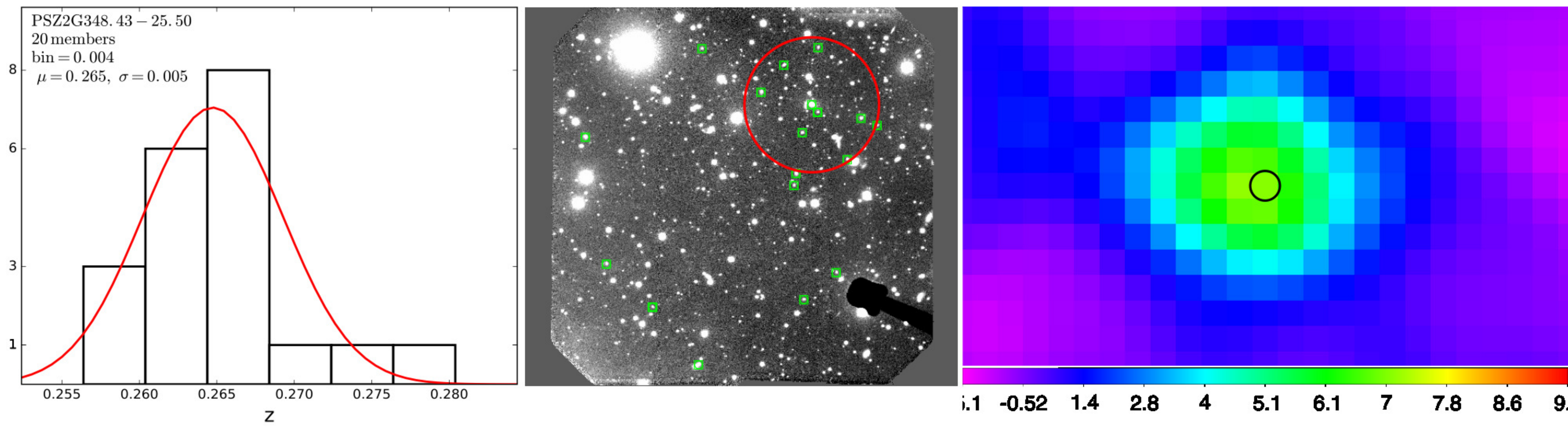}
\plotone{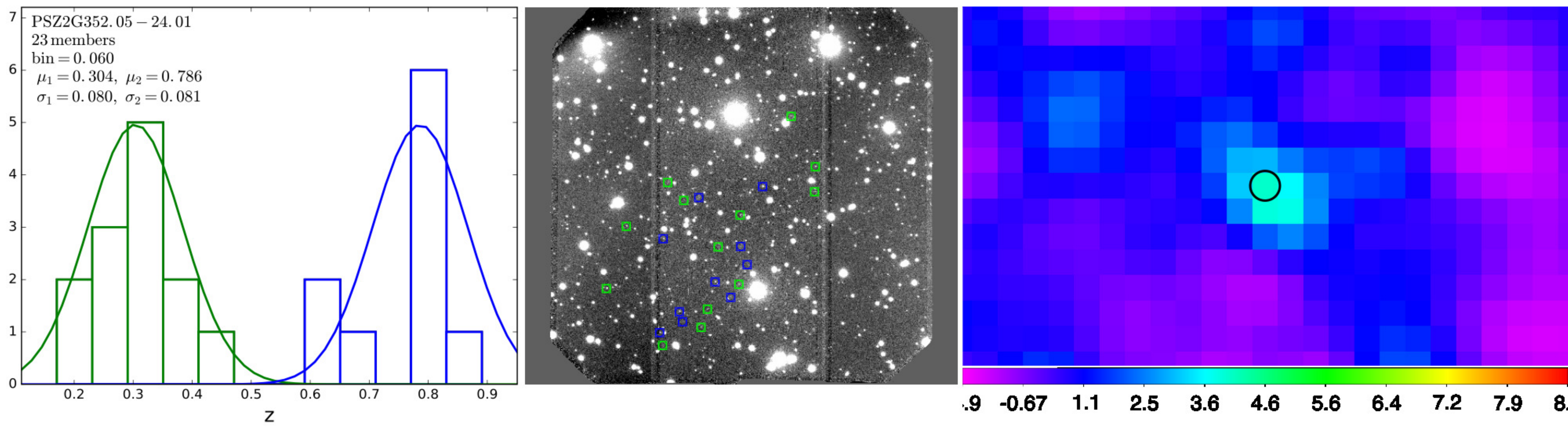}
\caption{Redshift histograms, optical images and SZ maps of clusters in the Southern sample. Symbols are the same as for Figure \ref{fig:zhist}. For PSZ2 G352.05-24.01, we know only the coordinates of the X-ray center, marked with a red cross.}
\label{fig:zhist2}
\end{center}
\end{figure*}
\clearpage

\begin{figure*}[h!]
\begin{center}
\figurenum{B3}
\epsscale{1.2}
\plotone{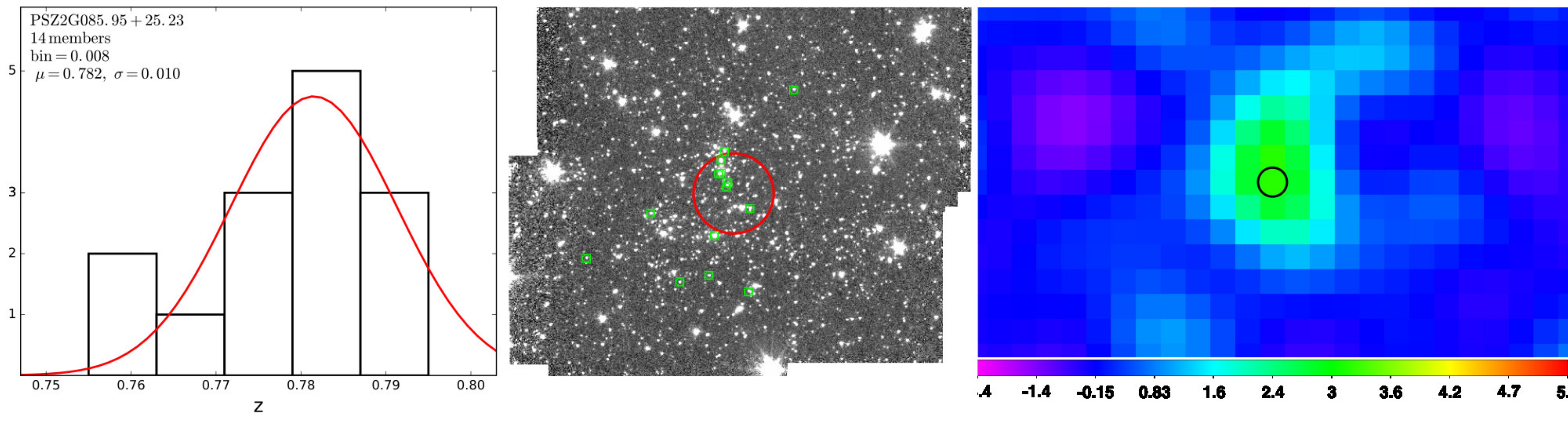}
\caption{Redshift histogram, IRAC image and SZ map of PSZ2 G085.95+25.23 observed at the Keck telescope. Symbols are the same as for Figure \ref{fig:zhist}.}
\label{fig:zhist3}
\end{center}
\end{figure*}

\section{Spectroscopic redshift catalog of PSZ2 G053.44-36.25 member galaxies}
\label{sec:appendix}

\begin{table}[h!]
\begin{center}
\caption{Catalog of galaxies detected for cluster PSZ2 G053.44-36.25. The full spectroscopic catalog is available in the online version of the journal.}
\label{tab:catalog}
\vspace{0.25cm}
\resizebox{!}{5.5cm}{
\begin{tabular}{llccccccccccccccc}
\tableline \tableline\\
ID & R.A.& Decl.&SPECZ&eSPECZ\\
     & (deg) & (deg) & & \\
\tableline \tableline \\
1&$21.58816$& $-1.08456$& $0.3306$& $0.0002$\\
2&$21.58506$& $-1.06186$& $0.3324$& $0.0001$\\
3&$21.58748$& $-1.05329$& $0.3363$&$0.0001$\\
4&$21.58530$& $-1.08879$& $0.3304$& $0.0003$\\
5&$21.58638$& $-1.05156$& $0.3361$& $0.0003$\\
6&$21.58564$& $-1.06893$& $0.3258$& $0.0002$\\
7&$21.58671$& $-1.05585$& $0.3359$& $0.0001$\\
8&$21.58600$& $-1.06488$& $0.3301$& $0.0002$\\
9&$21.58632$& $-1.02193$& $0.3344$& $0.0003$\\
10&$21.58714$& $-1.04561$& $0.3277$& $0.0002$\\
11&$21.58603$& $-1.02659$& $0.3345$& $0.0002$\\
12&$21.58648$& $-1.05931$& $0.3239$& $0.0002$\\
13&$21.58509$& $-1.07221$& $0.3316$& $0.0002$\\
14&$21.58678$& $-1.07722$& $0.3322$& $0.0003$\\
15&$21.58659$& $-1.03027$& $0.3250$& $0.0003$\\
16&$21.58745$& $-1.03873$& $0.3335$& $0.0001$\\
17&$21.58458$& $-1.04332$& $0.3307$& $0.0002$\\
18&$21.58804$& $-1.03449$& $0.3891$& $0.0006$\\
19&$21.58677$& $-1.02851$& $0.3424$& $0.0004$\\
20&$21.58674$& $-1.04831$& $0.3276$& $0.0002$\\
\tableline \tableline\\
\end{tabular}}
\end{center}
\end{table}


\listofchanges

\end{document}